\documentclass[twocolumn]{aastex701}
\usepackage{adjustbox}
\usepackage{graphicx}
\usepackage{subfigure}
\usepackage{xcolor,colortbl}

\shorttitle{VLA FRAMEx~III: Radio Emission Mechanisms}
\shortauthors{Sargent et al.}

\defcitealias{2021ApJ...906...88F}{FRAMEx~I}
\defcitealias{2022ApJ...927...18F}{FRAMEx~II}
\defcitealias{2022ApJ...936...76S}{FRAMEx~III}
\defcitealias{2023ApJ...958...61F}{FRAMEx~IV}
\defcitealias{2024ApJ...961..109S}{FRAMEx~V}
\defcitealias{2024ApJ...961..230S}{Paper~I}
\defcitealias{2025ApJ...986..194S}{Paper~II}

\graphicspath{{./}{figures/}}

\begin{document}

\title{VLA FRAMEx. III. Circumnuclear Radio Emission Mechanisms in Hard X-ray Selected Active Galactic Nuclei}

\AuthorCollaborationLimit=8

\author[0000-0002-8049-0905]{Andrew J.\ Sargent}
\email[show]{andrew.j.sargent2.civ@us.navy.mil}
\affiliation{United States Naval Observatory, 3450 Massachusetts Ave., NW, Washington, DC 20392, USA}
\affiliation{Department of Physics, The George Washington University, 725 21st St. NW, Washington, DC 20052, USA}

\author[0000-0001-9149-6707]{Alexander J.\ van der Horst}
\email{ajvanderhorst@email.gwu.edu}
\affiliation{Department of Physics, The George Washington University, 725 21st St. NW, Washington, DC 20052, USA}

\author[0000-0002-4146-1618]{Megan C. Johnson}
\email{mejohnso@nsf.gov}
\affiliation{National Science Foundation, 2415 Eisenhower Ave., Alexandria, VA 22314, USA}

\author[0000-0002-4902-8077]{Nathan J.\ Secrest}
\email{nathan.j.secrest.civ@us.navy.mil}
\affiliation{United States Naval Observatory, 3450 Massachusetts Ave., NW, Washington, DC 20392, USA}

\author[0000-0002-3365-8875]{Travis C.\ Fischer}
\email{tfischer@stsci.edu}
\affiliation{AURA for ESA, Space Telescope Science Institute, 3700 San Martin Drive, Baltimore, MD 21218, USA}

\author[0000-0003-4727-2209]{Onic I. Shuvo}
\email{oishuvo@umbc.edu}
\affiliation{Department of Physics, University of Maryland Baltimore County, 1000 Hilltop Circle, Baltimore, MD 21250, USA}

\author[0000-0002-1292-1451]{Macon A. Magno}
\email{macon.a.magno@tamu.edu}
\affiliation{George P. and Cynthia Woods Mitchell Institute for Fundamental Physics and Astronomy, Texas A\&M University, College Station, TX, 77845, USA}
\affiliation{CSIRO Space and Astronomy, ATNF, PO Box 1130, Bentley WA 6102, Australia}

\author[0000-0002-0819-3033]{Luis C. Fernandez}
\email{lfernan@gmu.edu}
\affiliation{Gunston Middle School, Arlington Public Schools, 2700 S. Lang Street Arlington, VA 22206, USA}

\begin{abstract}
We present Stokes I continuum analysis for a volume-limited sample ($<40~\rm{Mpc}$) of hard X-ray selected active galactic nuclei (AGNs) using $4-12$ GHz observations with the Karl G. Jansky Very Large Array (VLA). All of the 25 sources analyzed here have previously been observed with the Very Long Baseline Array (VLBA) to probe their subparsec projected physical scales, but detected emission has only been measured for 12 of the sources at $C$ band (4.4 GHz), despite expectations. We determined that coronal emission is unlikely to be a dominant emission mechanism for the sources not detected by the VLBA, and the emission measured with the VLA is likely produced beyond parsec spatial scales. We also explore potential radiation mechanisms for the circumnuclear radio emission that is produced beyond the observable ~parsec physical scales probed with the VLBA but within the $\leq30-110$ parsec spatial scales observed with the VLA. From an energetics perspective, we find that all targets have extranuclear radio emission that is compatible with AGN winds, assuming a maximum of 10\% of the bolometric output can supply the mechanical energy observed. We also find that the excess emission is likely too strong for star formation alone when compared to results from optical spectroscopy, but may contribute in smaller capacities.
\end{abstract}

\section{Introduction} \label{sec:intro}

Supermassive black holes (SMBHs, black holes with masses above $\sim10^6$ solar masses) are the central engines powering active galactic nuclei (AGNs) and quasars observed at all cosmic epochs. 
As SMBHs accrete nearby matter, they interact with their host environment through a mechanism known as AGN feedback, which is primarily described by two modes: kinetic (or mechanical) and radiative \citep{2012ARA&A..50..455F}. Kinetic-mode feedback is often associated with luminous radio jets, which can extend beyond the host galaxy and may be responsible for heating the intercluster medium in galaxy clusters on Mpc scales \citep{2016ApJ...829...90Y}. Conversely, radiative mode output uniquely changes the ionization state in the immediate vicinity ($\lesssim1$\,kpc) of the AGN. The coupling of radiation and cold gas may subsequently produce AGN winds which impact the surrounding interstellar medium at speeds on the order of $10^{3}~{\rm km~s^{-1}}$ \citep[for example, see][]{1984RvMP...56..255B,2017ApJ...834...30F,2020MNRAS.491.1518R,2025ApJ...984...32R}.
AGN winds can shock the host environment and accelerate electrons that interact with magnetic fields to produce synchrotron emission observed in the radio \citep[e.g.][]{2014MNRAS.442..784Z}.

The two modes of feedback are not necessarily mutually exclusive, and discerning the physical processes that drive AGN feedback has been an ongoing topic for decades. The difficulty of analyzing how the AGN is impacting its host environment is compounded when considering that only $5-10\%$ of all AGNs that are observable are considered to be ``radio-loud'' \citep{1984RvMP...56..255B,2009AJ....137...42R,2025ApJS..280...23A}.\footnote{Radio-loud AGNs are typically defined as the ratio of radio-to-optical luminosities being $>10$ \citep{1989AJ.....98.1195K}.} These objects appear to be dominated by bright radio jets that outshine their hosts and are observed at moderate redshifts ($z\sim1$). The extreme distances of the vast majority of radio-loud sources preclude a detailed analysis of jet-related activity near the accretion disk, which is only a few light-days in extent \citep{2022ApJ...929...19G}. For example, the accretion disk is well below the resolving capability of even very long baseline interferometry (VLBI) for a quasar at $z=1$, where 1 milliarcsecond (mas) corresponds to 8 parsec ($h=0.7$).  Therefore, observations of nearby AGNs are required in order to disambiguate feedback processes.

The nearer ``radio-quiet'' AGNs are without powerful jets and have radio luminosities that are approximately three orders of magnitude fainter than their radio-loud counterparts \citep[][]{2019NatAs...3..387P}. They contain compact radio cores that are sometimes accompanied by additional components or jet-like structures when observed at arcsecond resolution \citep[examples include][]{1984ApJ...285..439U,1995MNRAS.276.1262K,2001ApJS..133...77H,2019MNRAS.485.2710J,2020MNRAS.492.4216S}. The resolved radio structure in these AGNs often appears aligned with ionized emission and molecular outflows associated with nuclear activity, leading to the hypothesis that while the observed radio structure may appear jet-like, the radio emission is actually a byproduct of a radiative outflow shocking the host medium that is observed as projected emission onto the host disk \citep[e.g.][]{2023ApJ...953...87F,2022A&A...665L..11P,2024ApJ...977..156D}.

For unresolved radio emission regions coincident with the nucleus, the ubiquity of synchrotron radiation at $\sim$\,GHz frequencies and the variety of potential mechanisms from which they arise presents a challenge for discriminating between pure AGN processes and extranuclear sources \citep[for an excellent review of radio emission origins in AGNs, see][]{2019NatAs...3..387P}. From one perspective, the tight correlation between radio and X-ray luminosities in radio-quiet quasars suggests similarity to the magnetic reconnection activity observed in coronally active cool stars \citep[i.e. $L_{\rm R}/L_{\rm X}\sim10^{-5}$;][]{1993ApJ...405L..63G,2008MNRAS.390..847L}, potentially implying that the radio emission in AGNs is tied directly to the embedded accretion activity. Alternatively, the production of synchrotron radiation associated with the base of a highly collimated jet in the inner $\ll 1~\rm pc$ of an AGN may dominate the unresolved radio emission \citep[pertinent reviews on relativisitic AGN jets include][]{1984RvMP...56..255B,2019ARA&A..57..467B}. Another option for the source of the unresolved emission may be due to high-velocity AGN winds or uncollimated outflows which can interact with and shock the immediate host medium, as discussed at the beginning of this section. Radio observations also provide an excellent probe of recent star-formation activity which may further contaminate the unresolved structure. In the star-forming processes, electrons are accelerated through magnetic fields within supernova remnants which result from the collapse of massive stars \citep[see, e.g.,][]{1992ARA&A..30..575C,2001ApJ...554..803Y}. Thus, if the accretion processes near the SMBH are not energetically preeminent, the unresolved compact radio structure may be dominated by mechanisms produced beyond the regions with direct accretion activity. For example, though faint AGN activity (with 1.4 GHz radio peak flux densities of $37-59~\rm{\mu Jy~bm^{-1}}$) was recently detected in 4 out of 500 star-forming galaxies, the excess AGN emission measured from high-resolution VLBI observations appeared to have low-impact on the overall radio-infrared correlation despite non-negligible radio flux contributions \citep{2025arXiv250917536P}. As the radio-infrared correlation stems from star forming properties of the host galaxy, these results imply that the star formation relation is robust against a faint compact core.

To address the relationship between the radio properties of AGNs and their accretion states, we have been leading the Fundamental Reference AGN Monitoring Experiment \citep[FRAMEx;][]{2020jsrs.conf..165D}, a program to characterize the accretion-level dynamics from a volume-limited sample of AGNs. We initially probed the subparsec/mas emission regime using 6\,GHz observations with the Very Long Baseline Array (VLBA), but so far we have measured radio emission for only 12 out of 25 targets \citep[see ][hereafter \citetalias{2021ApJ...906...88F} and \citetalias{2022ApJ...936...76S}, respectively]{2021ApJ...906...88F,2022ApJ...936...76S}. We have subsequently reobserved all 25 \citetalias{2021ApJ...906...88F} targets with the Karl G. Jansky Very Large Array (VLA) to obtain a uniform data set, which we observed across a continuous $4-12$ GHz frequency band (\citealt{2025ApJ...986..194S}; hereafter \citetalias{2025ApJ...986..194S}). The lack of detections for many of the FRAMEx targets with VLBA measurements is in direct contrast to the 100\% detection rate at arcsecond resolution in \citetalias{2025ApJ...986..194S}, implying that much of the radio emission resolves out of VLBA observations as it is produced at larger, parsec spatial scales.

In this paper we analyze the radio emission observed by the VLA by modeling potential mechanisms in order to constrain its source. In Section \ref{sec:methodology} we briefly discuss the sample and analysis procedures used in our previous works, in Section \ref{sec:analysis} we explore various radio emission mechanisms as they relate to our observations, and we discuss and summarize our results in the broader context in Section \ref{sec:conclusion}.

\section{Methodology}
\label{sec:methodology}

The target sample in this paper consists of the original 25 AGNs described in \citetalias{2021ApJ...906...88F}. The sample was selected from the Neil Gehrels Swift Observatory ({\it Swift}) Burst Alert Telescope (BAT) 105-month all-sky hard X-ray survey \citep{2018ApJS..235....4O}. The targets selected for observation have a $14-195~{\rm keV}$ hard X-ray minimum luminosity of $10^{42}~{\rm erg~s^{-1}}$ and form a volume-complete sample of all AGNs within 40 Mpc and the declination limits chosen in \citetalias{2021ApJ...906...88F} ($-30^{\circ}<\delta<60^{\circ}$).

\label{sec:radio}
We obtained two observations for each target with the VLA in its most extended A-configuration: one spanned the full $4-8~{\rm GHz}$ bandwidth of the C-band receiver, while the other spanned the full $8-12~{\rm GHz}$ bandwidth of the X-band receiver. In both observations, we additionally acquired full Stokes polarization, the results of which will be discussed in future work. These observations allow for a uniform set of data at the highest achievable resolution that the VLA can provide at these frequencies. In \citet[][hereafter \citetalias{2024ApJ...961..230S}]{2024ApJ...961..230S}, we described the calibration and imaging techniques used for these observations and conducted a case study of the AGN in NGC~4388. In \citetalias{2025ApJ...986..194S}, we expanded the calibration to the full sample to produce the Stokes I broadband imaging results centered at 6 and 10 GHz. In that work we additionally measured the spectral energy distributions across the full $4-12$ GHz observing frequency range with a spectral resolution of 0.128 GHz for most sources to calculate spectral indices in the nuclear regions for each target. The analysis in this paper serves as a direct continuation of the results presented in \citetalias{2025ApJ...986..194S}.

In Table \ref{tab:framex}, we list the FRAMEx sample and the associated AGN type, distance, and black hole mass for all targets. We additionally supplement each target with their corresponding spectral indices from our VLA observations and their 4.4 GHz radio luminosities, 4.4 GHz radio luminosities from our VLBA observations \citepalias{2024ApJ...961..109S}, $2-10~{\rm keV}$ X-ray nuclear luminosities from our {\it Swift} X-ray Telescope (XRT) observations, and their bolometric luminosities. Note that no nuclear VLA luminosity was measured for the target NGC~1068 in \citetalias{2025ApJ...986..194S} as it is contaminated with additional sources near the nucleus at VLA resolution (see Figure 6 in \citetalias{2021ApJ...906...88F}), and thus no accurate flux measurement of the central $\lesssim30~{\rm pc}$ could be obtained. We instead use measurements from the literature which include higher resolution imaging captured from the combination of \textit{enhanced} Multi-Element Remotely Linked Interferometer Network (\textit{e}-MERLIN) and VLA observations \citep{2024MNRAS.52711756M,2025MNRAS.539..808M}. Throughout this paper, we quote uncertainty levels when available, and propagate the uncertainties for the measurements throughout this work.

\startlongtable
\tabletypesize{\footnotesize}
\begin{deluxetable*}{rcCCCCCCC}
\tablecaption{FRAMEx Sample}
\tablehead{
\colhead{Target}
& \colhead{Type}
& \colhead{Distance}
& \colhead{$\log M_{\rm BH}$}
& \colhead{$\alpha$}
& \colhead{$\log L^{\rm VLA}_{\rm C}$}
& \colhead{$\log L^{\rm VLBA}_{\rm C}$}
& \colhead{$\log L_{2-10~{\rm keV}}$}
& \colhead{$\log L_{\rm Bol.}$}
\\
& 
& \colhead{(Mpc)}
& \colhead{($M_{\odot}$)} 
&
& \colhead{($\rm erg~s^{-1}$)} 
& \colhead{($\rm erg~s^{-1}$)} 
& \colhead{($\rm erg~s^{-1}$)} 
& \colhead{($\rm erg~s^{-1}$)} 
}
\startdata
\multicolumn{9}{c}{VLBA Detection} \\
\hline
NGC~1052 &   Sy2 & 21.5 & 8.67 &   0.29 \pm 0.01        & 39.61\pm0.02          &  39.44\pm0.01^{\rm V}    & 41.62^{\rm c}           & 43.14^{+0.04}_{-0.05} \\
NGC~1068 & Sy1.9 & 16.3 & 6.93 &   0.00 \pm 0.1^{\rm a} & 37.47\pm0.01^{\rm b}  &  36.60\pm0.05^{\rm V}    & 42.93^{\rm c}           & 42.89^{+0.03}_{-0.05} \\
NGC~2110 &   Sy2 & 33.6 & 9.38 &   0.21 \pm 0.01        & 38.55\pm0.02          &  38.41\pm0.01^{\rm V}    & 43.01^{+0.05}_{-0.04}         & 44.54^{+0.00}_{-0.01} \\
NGC~2782 &   Sy1 & 36.6 & 6.07 &  -0.75 \pm 0.02        & 37.39\pm0.02          &  36.51\pm0.04^{\rm III}  & 41.30^{+0.5}_{-0.4}           & 43.19^{+0.07}_{-0.14} \\
NGC~2992 & Sy1.9 & 33.2 & 8.33 &  -0.39 \pm 0.01        & 37.80\pm0.02          &  37.03\pm0.03^{\rm V}    & 43.12^{+0.02}_{-0.02}         & 43.54^{+0.04}_{-0.05} \\
NGC~3079 &   Sy2 & 15.9 & 6.38 &  -0.20 \pm 0.01        & 38.35\pm0.02          &  38.12\pm0.01^{\rm V}    & 42.58                   & 42.23^{+0.05}_{-0.01} \\
NGC~4151 & Sy1.5 & 14.2 & 7.55 &  -0.51 \pm 0.01        & 37.86\pm0.02          &  37.48\pm0.04^{\rm V}    & 42.78^{+0.06}_{-0.08}         & 44.04^{+0.00}_{-0.00} \\
NGC~4235 & Sy1.2 & 34.5 & 7.55 &   0.54 \pm 0.01        & 37.48\pm0.02          &  37.30\pm0.04^{\rm V}    & 42.02^{+0.06}_{-0.06}         & 43.64^{+0.03}_{-0.03} \\
NGC~4388 &   Sy2 & 16.8 & 6.94 &  -0.32 \pm 0.02        & 36.76\pm0.02          &  35.94\pm0.06^{\rm V}    & 43.02^{+ 0.2}_{ -0.1}         & 43.86^{+0.01}_{-0.01} \\
NGC~4593 &   Sy1 & 38.8 & 6.88 &  -0.15 \pm 0.04        & 37.02\pm0.03          &  36.79\pm0.05^{\rm V}    & 42.81^{+0.03}_{-0.03}         & 44.05^{+0.03}_{-0.01} \\
NGC~5290 &   Sy2 & 37.1 & 7.78 &  -0.14 \pm 0.01        & 37.70\pm0.02          &  37.54\pm0.01^{\rm V}    & 42.39^{+0.05}_{-0.05}         & 43.28^{+0.05}_{-0.08} \\
NGC~5506 & Sy1.9 & 26.7 & 6.96 &  -0.80 \pm 0.01        & 38.74\pm0.02          &  38.46\pm0.01^{\rm V}    & 42.96^{+ 0.1}_{-0.03}         & 44.13^{+0.01}_{-0.00} \\
\hline
\multicolumn{9}{c}{VLBA non-detection} \\
\hline
NGC~1320 &           Sy2 & 38.4 & 7.96 & -0.65 \pm 0.04  & 37.08\pm0.03      & <35.92^{\rm III}  & \nodata                       & 43.28^{+0.07}_{-0.12} \\
 IC~2461 &           Sy2 & 32.3 & 7.27 & -0.47 \pm 0.07  & 36.55\pm0.03      & <36.11^{\rm I}    & 41.79^{\rm c}           & 43.28^{+0.01}_{-0.03} \\
NGC~3081 &           Sy2 & 34.5 & 7.74 & -0.45 \pm 0.05  & 36.86\pm0.03      & <35.76^{\rm III}  & 42.65^{+ 0.1}_{ -0.2}         & 43.51^{+0.01}_{-0.03} \\
NGC~3089 & Sy2 candidate & 38.8 & 6.55 & -0.71 \pm 0.06  & 36.44\pm0.02      & <35.95^{\rm III}  & 41.56^{+ 0.2}_{ -0.1}         & 44.03^{+0.07}_{-0.25} \\
NGC~3227 &         Sy1.5 & 16.8 & 6.77 & -0.92 \pm 0.01  & 37.35\pm0.02      & <35.68^{\rm I}    & 42.43^{+0.03}_{-0.03}         & 43.44^{+0.01}_{-0.01} \\
NGC~3786 &         Sy1.9 & 38.4 & 7.48 & -0.78 \pm 0.05  & 37.05\pm0.11      & <36.40^{\rm I}    & 42.12^{\rm c}           & 43.31^{+0.06}_{-0.10} \\
NGC~4180 &           Sy2 & 30.1 & 7.63 & -0.63 \pm 0.05  & 36.54\pm0.02      & <36.39^{\rm I}    & 41.94                   & 43.09^{+0.07}_{-0.14} \\
NGC~5899 &           Sy2 & 37.1 & 7.69 & -0.86 \pm 0.02  & 37.43\pm0.02      & <36.31^{\rm I}    & 42.12^{+ 0.1}_{ -0.1}         & 43.43^{+0.04}_{-0.05} \\
NGC~6814 &         Sy1.5 & 22.4 & 7.04 & -0.61 \pm 0.09  & 36.54\pm0.04      & <35.56^{\rm III}  & 42.10^{+0.03}_{-0.03}         & 43.42^{+0.05}_{-0.01} \\
NGC~7314 &         Sy1.9 & 20.6 & 6.76 & -0.96 \pm 0.03  & 36.67\pm0.02      & <35.46^{\rm III}  & 42.02^{+0.03}_{-0.03}         & 43.27^{+0.05}_{-0.02} \\
NGC~7378 &           Sy2 & 37.1 & 4.93 & -0.67 \pm 0.14  & 36.04\pm0.06      & <36.35^{\rm I}    & 41.12^{+ 1.1}_{ -0.2}         & 43.26^{+0.08}_{-0.17} \\
NGC~7465 &           Sy2 & 28.4 & 6.54 & -0.94 \pm 0.04  & 36.79\pm0.03      & <35.62^{\rm III}  & 42.03^{+0.05}_{-0.05}         & 43.17^{+0.06}_{-0.08} \\
NGC~7479 &           Sy2 & 34.0 & 7.61 & -0.39 \pm 0.01  & 37.22\pm0.02      & <36.25^{\rm I}    & 44.34                   & 43.27^{+0.06}_{-0.10}
\enddata
\tablecomments{Column (1) Target name. Column (2) AGN type according to \cite{2025ApJ...989..161L}. Column (3) Distance to target. Column (4) Log of black hole mass. Column (5) Spectral indices of nuclei \citepalias[$S\propto\nu^{\alpha}$;][]{2025ApJ...986..194S}. Column (6) Radio luminosity determined from VLA observations at 4.4 GHz \citepalias{2025ApJ...986..194S}. Column (7) Radio luminosity determined from VLBA observations. Upper limits on the VLBA non-detections were determined from  observations as $5\times{\rm rms}$. Column (8) X-ray luminosity determined from \textit{Swift}/XRT observations \citep{2021ApJ...906...88F}, except for NGC~1052, NGC~1068, NGC~1320, IC~2461, and NGC~3786. Column (9) Log of bolometric luminosity using \textit{Swift}/BAT  $F_{14-195~{\rm keV}}$ measurements \citepalias[Table 7 in ][]{2021ApJ...906...88F}, where $L_{\rm bol}=8\times L_{14-195~{\rm keV}}$ \citep{2023MNRAS.518.2938T}.\\
$^I$ VLBA flux densities determined from \citetalias{2021ApJ...906...88F} (6 GHz)\\
$^{III}$ VLBA flux densities determined from \citetalias{2022ApJ...936...76S} (6 GHz)\\
$^{V}$ VLBA flux densities determined from \citetalias{2024ApJ...961..109S} (4.4 GHz)\\
$^{\rm a}$ \cite{2024MNRAS.52711756M} \\
$^{\rm b}$ Luminosity determined from 4.8 GHz \textit{e}-MERLIN observation \citep{2025MNRAS.539..808M} \\
$^{\rm c}$ \textit{Swift}/XRT $2-10$ keV intrinsic fluxes determined from Table 12 in \cite{2017ApJS..233...17R}
}
\label{tab:framex}
\end{deluxetable*}

\section{Analysis} \label{sec:analysis}

The FRAMEx project has revealed an inconsistency in radio flux when nearby AGNs are observed with the differing resolving capabilities of the VLBA versus the VLA. This discrepancy implies that a significant portion of the nuclear emission is produced within an unprobed region corresponding to distances of $\sim1-100~{\rm pc}$ from the SMBH (see Table \ref{tab:diagnostics} for size scales probed by each telescope for all targets). Only 12 out of 25 targets have been detected by the VLBA even after several targets were reobserved with deeper 4-hour observations that doubled the initial sensitivity \citepalias[see][]{2022ApJ...936...76S}, suggesting that these sources may lack any effective small-scale radio emission and may in fact be ``radio silent'' at parsec physical scales \citepalias[see][]{2021ApJ...906...88F}. In contrast, 100\% of the FRAMEx targets have radio emission detected above the $5\times\rm{rms}$ level with the VLA, and our full sample results in \citetalias{2025ApJ...986..194S} revealed that at VLA resolution the AGNs consist of a diverse assortment of morphologies, have a wide range of luminosities, and have a variety of spectral shapes. Therefore in this section we attempt to interpret the extranuclear radio emission mechanisms by analyzing possible sources of radiation measured in our VLA observations. We compare our VLA results from \citetalias{2025ApJ...986..194S} to the high-resolution radio observations with the VLBA and the X-ray observations conducted with \textit{Swift}, to constrain what the source of the radio emission is and how it 
corresponds to the tens of parsecs physically probed by the VLA (see Table \ref{tab:diagnostics}). We provide initial diagnostics for the AGNs in Table \ref{tab:diagnostics}, including the  ``excess'' emission observed by the VLA that becomes resolved away at VLBA resolution, along with ratios for the VLBA-to-VLA and radio-to-X-ray flux levels. In the subsections that follow ,we use these results as we step through several potential radiation origins and mechanisms, starting nearest the black hole and extending outward.

\begin{deluxetable*}{rCCCCCCC}
\tablecaption{AGN Diagnostics}
\label{tab:diagnostics}
\tablehead{
\colhead{Target} &
\colhead{$S_{\rm excess}$} &
\colhead{$S_{\rm VLBA}/S_{\rm VLA}$} &
\colhead{$\log{R^{\rm VLA}_{\rm X}}$} &
\colhead{$\log{R^{\rm VLBA}_{\rm X}}$} &
\colhead{${d_{\rm VLA}}$} &
\colhead{${d_{\rm VLBA}}$} &
\colhead{$R_{\rm RS}$}
\\
&
\colhead{(${\rm mJy~bm^{-1}}$)} &
\colhead{(\%)} & & &
\colhead{(${\rm pc}$)} &
\colhead{($10^{-3}~{\rm pc}$)} &
\colhead{($10^{-3}~{\rm pc}$)} 
}
\startdata
\multicolumn{8}{c}{VLBA Detection} \\
\hline
NGC~1052 & 530    \pm   90    & 68    \pm  3 & -2.02                 & -2.18                 & 76         &   666     & 53  ^{+5 }_{-5 }    \\
NGC~1068 & 17     \pm   1     & 13     \pm 1 & -5.50                 & -6.37                 &  5^{\rm a} &  471 &  6.8^{+0.5}_{-0.6}  \\ 
NGC~2110 &  16    \pm   3     & 72    \pm  4 & -4.47^{+0.05}_{-0.05} & -4.61^{+0.05}_{-0.04} & 94         &  1030     & 28  ^{+0.3}_{-0.4}  \\
NGC~2782 & 2.2    \pm   0.1   & 13    \pm  1 & -3.97^{+0.50}_{-0.40} & -4.80^{+0.50}_{-0.41} & 97         &   554     &  6.7^{+1.1}_{-1.8}  \\
NGC~2992 &  9.1   \pm    0.6  & 17    \pm  1 & -5.32^{+0.03}_{-0.03} & -6.09^{+0.04}_{-0.04} & 08         &  1024     & 11  ^{+1 }_{-1 }    \\
NGC~3079 &   68   \pm      9  & 59    \pm  2 & -4.24                 & -4.46                 & 46         &   358     & 16  ^{+2}_{-0.3}    \\
NGC~4151 &    40  \pm      4  & 42    \pm  1 & -4.92^{+0.06}_{-0.08} & -5.30^{+0.07}_{-0.09} & 32         &   330     & 13  ^{+0.1}_{-0.1}  \\
NGC~4235 & 1.6    \pm    0.4  & 67    \pm  6 & -4.55^{+0.06}_{-0.06} & -4.72^{+0.07}_{-0.07} & 83         &  1219     &  8.4^{+0.6}_{-0.6}  \\
NGC~4388 & 3.3    \pm    0.2  & 15    \pm  1 & -6.27^{+0.20}_{-0.10} & -7.09^{+0.20}_{-0.12} & 44         &   523     &  4.6^{+0.1}_{-0.1}  \\
NGC~4593 & 0.55   \pm   0.12  & 58    \pm  6 & -5.78^{+0.04}_{-0.04} & -6.02^{+0.05}_{-0.06} & 75         &  1153     &  6.1^{+0.5}_{-0.2}  \\
NGC~5290 & 2.1    \pm    0.4  & 69    \pm  3 & -4.68^{+0.05}_{-0.06} & -4.84^{+0.05}_{-0.05} & 82         &   761     &  9.6^{+1.3}_{-1.5}  \\
NGC~5506 & 70     \pm      8  & 52    \pm  2 & -4.22^{+0.10}_{-0.04} & -4.50^{+0.10}_{-0.03} & 85         &   781     & 30  ^{+0.9}_{-0.3}  \\
\hline
\multicolumn{8}{c}{VLBA Non-detection} \\
\hline
NGC~1320 &  1.5   \pm   0.1   & <5           & \nodata               &   \nodata &  100                &   \nodata &  5.4 ^{+0.9}_{-1.3} \\
 IC~2461 & 0.64   \pm   0.05  & <28          & -5.24                 & < -5.67   &   72                &   \nodata &  3.3 ^{+0.1}_{-0.2} \\
NGC~3081 &  1.2   \pm    0.1  & <7           & -5.79^{+0.10}_{-0.20} & < -6.90   &  115                &   \nodata &  5.1 ^{+0.2}_{-0.3} \\
NGC~3089 & 0.35   \pm   0.02  & <29          & -5.11^{+0.20}_{-0.10} & < -5.61   &  114                &   \nodata &  2.8 ^{+0.5}_{-1.2} \\
NGC~3227 &   15   \pm      1  & <2           & -5.08^{+0.04}_{-0.04} & < -6.75   &   42                &   \nodata &  7.2 ^{+0.3}_{-0.2} \\
NGC~3786 &  1.4   \pm    0.4  & <17          & -5.10                 & < -5.72   &   63                &   \nodata &  5.1 ^{+0.7}_{-1.0} \\
NGC~4180 & 0.73   \pm   0.04  & <52          & -5.39                 & < -5.55   &   75                &   \nodata &  3.4 ^{+0.6}_{-0.9} \\
NGC~5899 & 3.8    \pm    0.2  & <6           & -4.68^{+0.10}_{-0.10} & < -5.81   &   87                &   \nodata &  7.7 ^{+0.7}_{-0.9} \\
NGC~6814 & 1.3    \pm    0.1  & <10          & -5.56^{+0.05}_{-0.05} & < -6.54   &   75                &   \nodata &  3.4 ^{+0.4}_{-0.1} \\
NGC~7314 & 2.1    \pm    0.1  & <7           & -5.34^{+0.04}_{-0.04} & < -6.55   &   65                &   \nodata &  3.7 ^{+0.4}_{-0.2} \\
NGC~7378 & 0.15   \pm   0.02  & <100         & -5.08^{+1.10}_{-0.22} & < -4.76   &   77                &   \nodata &  2.0 ^{+0.4}_{-0.7} \\
NGC~7465 & 1.4    \pm    0.1  & <15          & -5.25^{+0.06}_{-0.06} & < -6.42   &   70                &   \nodata &  4.0 ^{+0.5}_{-0.7} \\
NGC~7479 & 2.8    \pm    0.1  & <8           & -7.12                 & < -8.09   &   75                &   \nodata &  6.1 ^{+0.9}_{-1.2} 
\enddata
\tablecomments{Column (1) Target name. Column (2) Excess flux density between the VLBA and VLA flux densities at 4.4 GHz. Column (3) VLBA flux density as a percentage of VLA flux density. VLBA-non detections have upper limits indicated (See Table \ref{tab:framex}). Column (4) Logarithm of VLA radio luminosity as a fraction of the X-ray luminosity (from Table \ref{tab:framex}). Column (5) Logarithm of VLBA radio luminosity as a fraction of the X-ray luminosity using luminosities in Table \ref{tab:framex} . Columns (6) and (7) Projected physical sizes for a point-source from the VLA and VLBA observations, respectively. Converted from the mean of the major and minor axis of the restoring beam, assuming that the nuclear source is an unresolved point source. Column (8) Theoretical size of radio emitting sphere \citep{2008MNRAS.390..847L}.\\
$^{\rm a}$ Determined using a restoring beamwidth of $0.06''$ \citep{2025MNRAS.539..808M}}
\end{deluxetable*}

\subsection{Coronal Emission Mechanisms}
\subsubsection{Origins}

Hard X-ray production occurs very close to the SMBH as is evidenced by the rapid ($\lesssim\rm{days}$) to long ($\sim\rm{years}$) variability timescales observed in AGNs \citep[implying light travel distances of $\sim10^{-3}-0.3$ pc;][]{1997MNRAS.292..679L,2004MNRAS.349.1435M,2006Natur.444..730M,2023MNRAS.526.1687T}. Much of the X-ray emission is assumed to originate from hot electrons in a corona above the mechanical accretion flow \citep{2018MNRAS.480.1819R}, and the region is expected to extend out to a few gravitational radii above the accretion disk \citep{2013ApJ...769L...7R,2016ApJ...821L...1N,2020A&A...644A.132U}. Radio luminosity ($L_{\rm R}$) at 5 GHz has been found by \cite{2008MNRAS.390..847L} to be strongly associated with its X-ray luminosity ($L_{\rm X}$, $0.2-20$ keV) for radio-quiet quasars, scaling as $L_{\rm R}/L_{\rm X}\sim10^{-5}$, potentially indicating that the two are intricately related and giving rise to the possibility that at least some radio emission is coupled with coronal activity.

The \cite{2008MNRAS.390..847L} work noted that the radio--X-ray luminosity scale factor has striking similarity with the Güdel-Benz relation \citep{1993ApJ...405L..63G} for coronally active stars, which extends 15 dex when AGNs are included, and they suggest that similar physical processes are at play in the two types of objects. Coronally active stars show agreement with solar flare emission at their peaks, perhaps indicating that the AGN coronal activity has a solar-like production mechanism of flares and likewise occur via magnetic reconnection events \citep{1994A&A...285..621B,2002ARA&A..40..217G}. In these events, thermal radiation from free-free emission in hot plasma ($T\sim10^{7}~\rm{K}$) emits soft X~rays, while non-thermal electrons in the magnetic fields are responsible for the radio emission through gyro-synchrotron radiation \citep{1993ApJ...405L..63G,2008MNRAS.390..847L,2022ApJ...926L..30V}.

This may lead to an expectation of variability in radio emission induced by the coronal activity in AGNs. The key point in this work is that radio and X-ray variability should be correlated. However, distinguishing variability for small-scale radiation mechanisms is not simple. While short-term variability ($\sim\rm{days}$) has been observed at VLA resolution for samples of nearby AGNs and may be intrinsic to the radio source, it is also consistent with extrinsic interstellar scintillation \citep{2005ApJ...618..108B,2005ApJ...627..674A}. One point of caution should be noted with respect to the $L_{\rm R}/L_{\rm X}$ relation observed by \cite{2008MNRAS.390..847L}. Their radio luminosities were determined from a sample of radio-quiet quasars observed with the VLA in its A-configuration \citep{1989AJ.....98.1195K,1994AJ....108.1163K} and the sample has a mean redshift of $z=0.17$ corresponding to kiloparsec scale emission. As we observed with our \citetalias{2021ApJ...906...88F} results, we found that a significant fraction of the radio flux density resolved away or is altogether missing in the subparsec nuclear regime (see $S_{\rm VLBA}/S_{\rm VLA}$ in Table \ref{tab:diagnostics}), an effect that may similarly affect the radio-quiet quasars. 

\begin{figure*}[ht!]
    \centering
    \subfigure[Radio vs. X-ray Luminosity (VLA Observations)][VLA vs. \textit{Swift}/XRT Luminosities]{
        \label{fig:lrlx_vla}\includegraphics[width=\columnwidth]{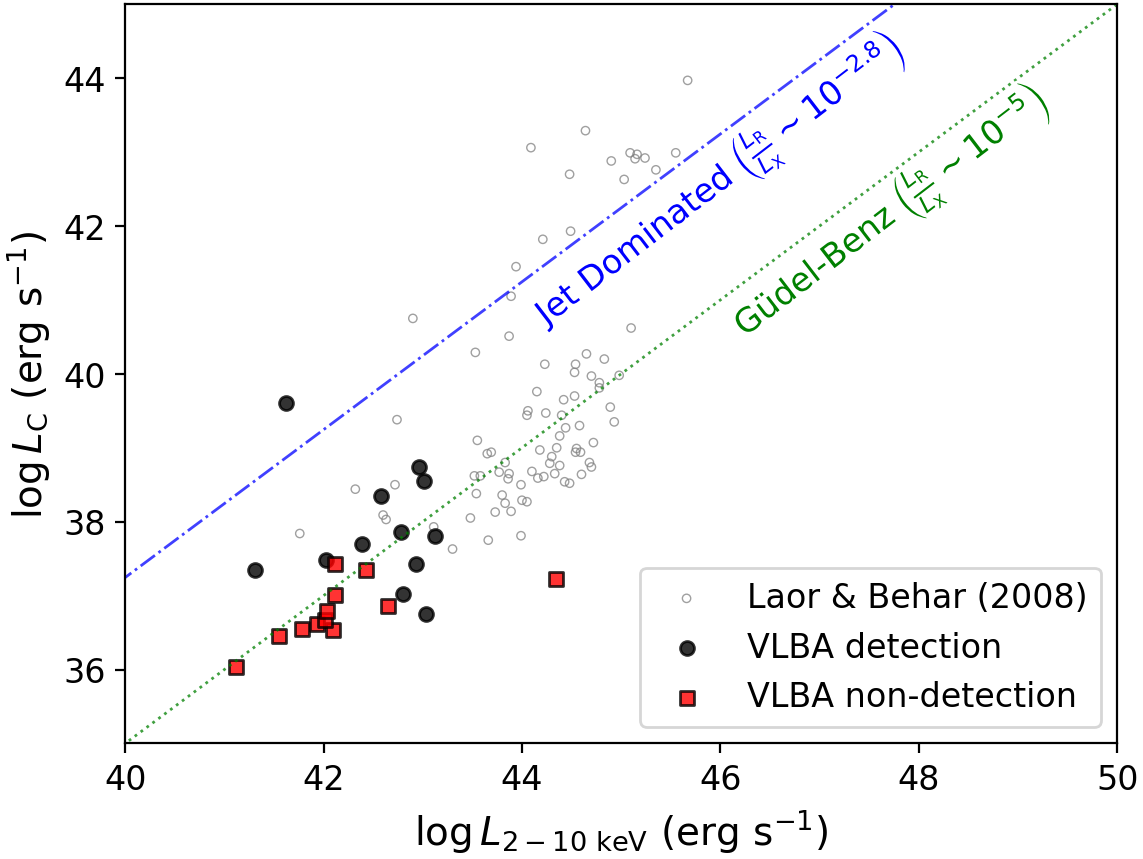}
    }
    \subfigure[Radio vs. X-ray Luminosity (VLBA Observations)][VLBA vs. \textit{Swift}/XRT Luminosities]{
        \label{fig:lrlx_vlba}\includegraphics[width=\columnwidth]{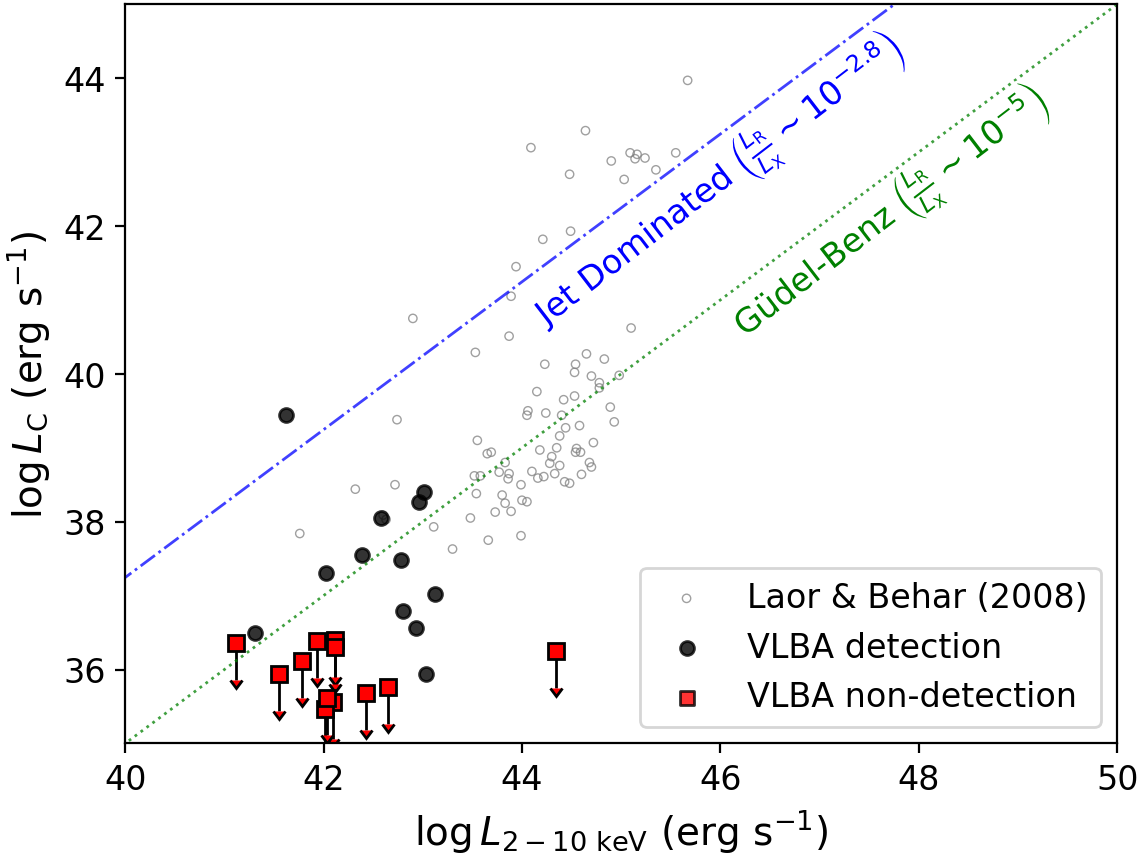}
    }
    \caption{\subref{fig:lrlx_vla} Radio luminosities from VLA observations \citepalias{2025ApJ...986..194S} versus X-ray luminosities \citepalias[][and this work]{2021ApJ...906...88F}. Solid black circles represent VLBA-detected objects and solid red squares represent VLBA non-detections. Hollow grey circles represent optically selected AGNs \citep{2008MNRAS.390..847L}. The dotted green line represents the Güdel-Benz relation for coronally active stars, and the dashed-dotted line represents the radio-loud AGN threshold defined by \cite{2007A&A...467..519P}. \subref{fig:lrlx_vlba} Same as Figure \ref{fig:lrlx_vla} but radio luminosities are determined from the VLBA measurements with upper limits indicated for VLBA non-detections ($5\times{\rm rms}$ at 6 GHz; see Table \ref{tab:framex}).}
    \label{fig:lrlx}
\end{figure*}

In Table \ref{tab:diagnostics}, we list the $L_{\rm R}/L_{\rm X}$ ratios of the FRAMEx sample, both for the VLA and the VLBA results using the 4.4 GHz radio luminosities and $2-10~{\rm keV}$ X-ray luminosities listed in Table \ref{tab:framex}. Several targets did not have $2-10~{\rm keV}$ X-ray observations in \citetalias{2021ApJ...906...88F} (NGC~1052, NGC~1068, and NGC~1320 were visibility constrained, the observation for IC~2461 was approved but not executed, and NGC~3786 did not have usable data), so we used the $2-10~{\rm keV}$ intrinsic fluxes from Table 12 in \cite{2017ApJS..233...17R} which were carried out with \textit{XMM-Newton} EPIC/PN observations.
We plot $L_{\rm R}$ versus $L_{\rm X}$ in Figure \ref{fig:lrlx} with radio luminosities using VLA measurements in panel \subref{fig:lrlx_vla} and VLBA measurements in panel \subref{fig:lrlx_vlba}, and we find that the FRAMEx sample appears to follow the Güdel-Benz relation at VLA resolution with a median value of $\log L_{\rm R}/L_{\rm X}=-5.1$ with a scatter of $0.7$ dex (without NGC~1052, which is three orders of magnitude brighter), potentially indicating coronal emission mechanisms. The VLBA resolution measurements appear to fall slightly above the Güdel-Benz relation and have a median value of $\log L_{\rm R}/L_{\rm X}=-5.7$ with a scatter of $0.7$ dex (for detected sources, excluding NGC~1052). Inferring coronal emission mechanisms based on the Güdel-Benz relation alone assumes that the radio and X-ray luminosities are directly correlated. We again want to emphasize the vastly different size scales probed with the \cite{2008MNRAS.390..847L} measurements ($\sim$kiloparsec) compared to our VLA (tens of parsecs) or VLBA (subparsec) results, and these initial comparisons alone should be heeded with caution.

\subsubsection{Estimating the Size of the Radio Emitter}
Estimating the physical sizes of the radio sources of our FRAMEx sample may provide insight on whether or not the radio emission has coronal origins. Size scales expected for a radio sphere range from being a few gravitational radii, comparable to that of the X-ray corona, out to $\sim10^4$ gravitational radii \citep{2016MNRAS.459.2082R}, which would extend into the narrow-line region \citep{2003ApJ...597..768S}. This corresponds to physical sizes between $3\times10^{-5} -3~\rm{pc}$ ($0.04-4\times10^3~{\rm lightday}$) from simulations by \cite{2016MNRAS.459.2082R} for a flat-spectrum optically-thick corona.

We can estimate the size of a theoretical compact radio sphere, $R_{\rm RS}$, which is a synchrotron source of emission with a uniform magnetic field:
\begin{equation}
R_{\rm RS}=0.47L_{30}^{0.4}L_{46}^{0.1}\nu_{\rm GHz}^{-1}~{\rm pc},
\end{equation}
where $L_{30}$ is the radio luminosity in units of $10^{30}~\rm{erg~s^{-1}~Hz^{-1}}$, $L_{46}$ is the bolometric luminosity in units of $10^{46}~\rm{erg~s^{-1}}$, and $\nu_{\rm GHz}$ is the frequency in units of GHz \citep{2008MNRAS.390..847L}. The estimation assumes equipartition between the magnetic and photon energy densities, which may be valid when they operate on similar time and length scales if there is a single source for the cosmic ray electrons and the magnetic field \citep{2019Galax...7...45S}. Using this interpretation, we find that the radio spheres range from $2-53\times10^{-3}~{\rm pc}$ (with a median of $6.1\times10^{-3}~{\rm pc}$), or $2-64~{\rm lightday}$ (median: 7.3 lightday), within the range of the simulations by \cite{2016MNRAS.459.2082R}. We note that these results are approximately in the ranges expected for the broad line region extents \citep[BLR, $\sim10-100$ lightday;][]{2005ApJ...629...61K} or the furthest extent of the accretion disk (\citealt{2022ApJ...929...19G} estimate $\sim3-5$ lightdays for AGNs).

In terms of the spectral structure, \cite{2016MNRAS.459.2082R} expect an optically thick coronal emitter to be a collection of overlapping synchrotron sources building up to a turnover at $\sim\rm GHz$ frequencies at the largest spatial extent. A flat spectrum above the turnover is expected in the $1-1000~{\rm GHz}$ frequency range followed by a downward turnover from an optically thin junction coming from near the base of the corona observed at hundreds of GHz. Many FRAMEx targets have a steep spectrum at VLA resolution at our $4-12$ GHz frequencies \citepalias[that is, $\alpha<-0.5$ for $S\propto\nu^{\alpha}$. See][]{2025ApJ...986..194S}, particularly the VLBA nondetections, which may indicate that the radio emission is not coronal. Two targets with VLBA detections have steep spectra: NGC~2782 ($\alpha=-0.75\pm0.02$) and NGC~5506 ($\alpha=-0.80\pm0.01$), though NGC~4151 ($\alpha=-0.51\pm0.01$) and NGC~2992 ($\alpha=-0.39\pm0.01$) straddle the flat spectrum line, so the excess radio emission detected for these targets indicate that it may not be due to coronal activity based on their spectra. Assuming that the coronal emitter dominates the radio spectrum, the most likely coronally active candidates to match the simulations are NGC~1052, NGC~2110, NGC~4235, which are all VLBA-detected FRAMEx targets with inverted spectra ($\alpha>0$) at VLA resolution.

The remaining (NGC~3079, NGC~4388, NGC~4593, and NGC~5290) have flat spectra (all with $\alpha\geq-0.32$), though they all still have a negative spectrum so the emission may not as likely be due to coronal activity.

\subsubsection{Extended Structure}
From a morphological perspective, \cite{2016MNRAS.459.2082R} expect coronally active AGNs to show resolved radio emission at VLBI resolution for nearby AGNs. This implies that VLBA non-detections likely lack any coronal activity, further strengthening the steep spectrum argument in the previous subsection. For the VLBA-detected targets, Figure 5 in \citetalias{2021ApJ...906...88F} shows that sources with more extended structure at VLA resolution are further displaced from the fundamental plane of black hole activity, a relation between the X-ray and radio luminosity combined with the black hole mass \citep{2003MNRAS.345.1057M}. Though our work in \citetalias{2021ApJ...906...88F} suggests that the fundamental plane breaks down at subparsec spatial scales, perhaps the sources with inverted VLA spectra that are the most pointlike at $\sim$arcsecond resolution, and thereby least displaced from the fundamental plane, are the FRAMEx targets most likely to be associated with coronal emission. Interestingly, our imaging results in \citetalias{2025ApJ...986..194S} show that the non-detections are primarily pointlike sources.
A full study on the morphology of FRAMEx targets is worth pursuing to examine how closely large-scale extended structure is related to nuclear emission mechanisms.

\subsubsection{Variability}

\begin{deluxetable}{cCCCc}
\label{tab:bayesvar}
\tablecaption{Bayesian X-ray Variability Analysis Results}
\tablehead{\colhead{Target} & \colhead{Variability} & \colhead{Credible} &  \colhead{$\log_{10} K$}  & Strength of evidence \\ & \colhead{Amplitude} & \colhead{interval} & & \\ & $10^{-4}$ & $10^{-4}$ & & }
\startdata
\multicolumn{5}{c}{Detections} \\
\hline
NGC~1052 &    0.63 & [  0.13,   2.82] &    -1.1 & Strong non-variable \\
NGC~1068 &    6.0  & [  4.8 ,   7.3 ] &     22  & Decisive variable \\
NGC~2110 &   53    & [ 48   ,  59   ] &  3492   & Decisive variable \\
NGC~2782 &    0.35 & [  0.05,   2.4 ] &    -0.2 & Inconclusive \\
NGC~2992 &    6.4  & [  5.0 ,   7.9 ] &    14   & Decisive variable \\
NGC~3079 &    3.9  & [  2.8 ,   5.0 ] &     8.3 & Decisive variable \\
NGC~4151 &  163    & [149   , 179   ] & 74366   & Decisive variable \\
NGC~4235 &   19    & [ 17   ,  21   ] &   493   & Decisive variable \\
NGC~4388 &   21    & [ 19   ,  24   ] &   705   & Decisive variable \\
NGC~4593 &   14    & [ 12   ,  16   ] &   200   & Decisive variable \\
NGC~5290 &    1.1  & [  0.08,   2.9 ] &     0.4 & Inconclusive \\
NGC~5506 &    6.2  & [  4.8,    7.7 ] &    13   & Decisive variable \\
\hline
\multicolumn{5}{c}{Non-detections} \\
\hline
NGC~1320 &  0.27 & [ 0.04,  2.12] &  -0.3 & Inconclusive \\
 IC~2461 &  0.40 & [ 0.07,  2.26] &  -0.7 & Substantial non-variable \\
NGC~3081 &  5.4  & [ 3.8 ,  7.0 ] &   6.6 & Decisive variable \\
NGC~3089 &  0.23 & [ 0.05,  1.63] &  -0.8 & Substantial non-variable \\
NGC~3227 & 19    & [17   , 22   ] & 612   & Decisive variable \\
NGC~3786 &  0.15 & [ 0.02,  1.47] &  -0.1 & Inconclusive \\
NGC~4180 &  0.26 & [ 0.04,  1.95] &  -0.4 & Inconclusive \\
NGC~5899 &  0.96 & [ 0.12,  2.80] &  -0.4 & Inconclusive \\
NGC~6814 &  7.3  & [ 5.8 ,  8.9 ] &  18   & Decisive variable \\
NGC~7314 &  0.66 & [ 0.24,  2.35] &  -2.4 & Decisive non-variable \\
NGC~7378 &  0.20 & [ 0.06,  1.35] &  -1.3 & Strong non-variable \\
NGC~7465 &  4.3  & [ 2.5 ,  5.8 ] &   5.2 & Decisive variable \\
NGC~7479 &  0.33 & [ 0.08,  1.76] &  -1.4 & Strong non-variable 
\enddata
\end{deluxetable}

A potential indicator of whether the progenitor of the radio emission in AGNs is due to coronal processes \citep[the `smoking gun,' according to][]{2019NatAs...3..387P} is the observation of a radio flare preceding an X-ray flare, thereby replicating the Neupert Effect observed on the Sun and some active stars \citep{1968ApJ...153L..59N,2002ARA&A..40..217G}. In this scenario, the radio is the prompt measure of the coronal heating as electrons are injected into the flare, whereas the accumulation of energy in the hot plasma is indicated by the longer cooling timescales observed in X~rays \citep{2002ARA&A..40..217G,2008MNRAS.390..847L}. This effect leads to the direct proportionality of the light curves of the radio luminosity and that of the time derivative of the X-ray luminosity. 

Variability observations at the subparsec scale are essentially covered by the VLBA monitoring aspect of the FRAMEx program, and \citetalias{2022ApJ...927...18F} suggested coronal origins in the case of the subparsec radio emission in NGC~2992, where an anti-correlation between simultaneous VLBI and X-ray flux measurements was observed. Further candidates for coronal origins of radio emission may be targeted in AGN sources with observed X-ray variability. Hard X rays probe the innermost regions of AGNs and are representative of the intrinsic radiation attributes of AGN activity \citep{2014A&A...563A..57S,2023MNRAS.526.1687T}. The \textit{Swift}/BAT 157 Month AGN Survey \citep{2025ApJ...989..161L} contains $\sim200$ megaseconds of historical hard X-ray measurements and we have used it to conduct a variability analysis.

Using the Crab-weighted monthly lightcurve data, we compared the Bayesian evidence between a constant model, taking only the mean X-ray flux as a free parameter, and a variable model, taking also an intrinsic scatter term. The likelihoods for both models assume a normal (Gaussian) sampling distribution, with the variability model adding the intrinsic scatter term in quadrature to the formal measurement uncertainties. We used flat (uninformative) priors for both models. Because the models have low dimensionality (1 or 2), we evaluated their likelihoods on a grid. The Bayesian evidence is determined by calculating the fully marginalized likelihoods. We determined credible intervals on the mean X-ray rate and, for the variable model, intrinsic dispersion via posterior integration.

We present the statistical variability results in Table \ref{tab:bayesvar} and indicate the variability amplitude, its credible interval at the 5th and 95th percentiles, and the corresponding Bayes factor. We classify each target's strength of evidence utilizing the scale provided by \cite{Kass01061995}. We find that there are 9 variable sources and 3 non-variable (or inconclusive) sources for the VLBA detections, while the non-detections have 4 variable and 9 non-variable (or inconclusive) targets. We next quantify whether there is any statistical significance using Fisher's exact test, given the null hypothesis that there is no correlation between a FRAMEx target's detection status and whether that status implies variability (or not). We measure a $p$-value of 0.05, which marginally suggests that detected FRAMEx targets appear to be variable while FRAMEx non-detections appear to be non-variable, assuming that non-detections are not detectable at VLBA resolution.

The result of Fisher's exact test is more meaningful if there are no systematic differences between the detected and undetected sources, so we conducted a two-sample Komolgorov-Smirnov test on the target distances under the null hypothesis that there is no correlation between each sample and their respective distances. For this statistical test we measure a $p$-value of 0.66, indicating that we cannot conclude that the null hypothesis is rejected. We also analyzed the hard X-ray luminosities to determine if there are potentially any systematic differences causing one sample to have lower signal-to-noise \textit{Swift} BAT data. We calculated both the two-sample Kolmogorov-Smirnov test and the Anderson-Darling test and we measure a $p$-value of 0.06 for both tests. Neither test rejects the null hypothesis.

As there is no clear indication that there are statistically significant systematic differences between the two samples combined with the results of our X-ray variability analysis, it is possible that targets not detected by the VLBA are statistically less likely to be variable. This implies that there is no significant coronal emission component in the radio for the non-variable sources. The variable sources on the other hand appear more likely to be detected by the VLBA and further variability studies on the VLBA non-detections may warrant reobservation. However, three out of four of these targets (NGC~3081, NGC~6814, and NGC~7465) were already targeted in our VLBA deep observations \citepalias{2022ApJ...936...76S} and they were still not detected. This does not necessarily rule out radio variability for these sources but may indicate that they are too faint in the radio for detections at VLBA resolution.

\subsection{Jets \& Bubbles}

\subsubsection{Jetted AGN Hosts}
Radio emission in AGNs may be produced in the form of highly collimated jets when synchrotron-emitting electrons are relativistically accelerated along magnetized plasma \citep{1979ApJ...232...34B,1984RvMP...56..255B}. They often flow bilaterally out to hundreds of kiloparsecs from the nucleus of an AGN, perhaps as an exhaust channel for excess pressure \citep{1974MNRAS.169..395B}, and terminate with large-scale extended lobes such as the kiloparsec scale radio structures observed in one of the earliest known examples of radio jets, Cygnus~A \citep{1953Natur.172..996J,1974MNRAS.166..305h,1974MNRAS.169..395B}. 

However, not all active galaxies host prominent jets: they are frequently associated with radio-loud AGNs which are bright radio sources that comprise roughly $5-10$\% of all AGNs \citep[][and references therein]{1989AJ.....98.1195K}. Radio-loud AGNs are typically hosted within elliptical galaxies and are approximately three orders of magnitude brighter in the radio than their radio-quiet counterparts which make up the remaining $\sim90-95\%$ of AGNs and tend to have hosts with flattened disk-shaped morphologies \citep{2007ApJ...658..815S,2019NatAs...3..387P}. The radio-to-optical luminosity ($R_{\rm O}$)\footnote{$R_{\rm O}$ is typically defined as the ratio of luminosities at 6 cm radio wavelengths to $4400~\text{\AA}$ optical wavelengths \citep{1989AJ.....98.1195K}.} has historically been used to delineate radio-loud from radio-quiet AGNs as it has been observed to be bimodal, typically having a demarcation of $R_{\rm O}$ set at $10$ for whether the source is radio-loud \citep[][]{1980A&A....88L..12S,1989AJ.....98.1195K,1992ApJ...391..560V,1992ApJ...396..487S}. There are not necessarily two strictly distinct populations of AGNs, however, as several studies suggest that there is an intermediate population in a transition region \citep{2000ApJS..126..133W,2009AJ....137...42R}.

In this work, we use an alternative for determining radio loudness using 5 GHz radio and $2-10$ keV X-ray luminosities \cite[$R_{\rm X}$;][]{2003ApJ...583..145T}. Utilizing X-ray luminosities as the radio-loudness indicator avoids confusion with stellar processes that are typical in the optical band and, importantly, is measurable even for highly absorbed nuclei when extinction effects may increase the radio-loudness \citep{2001ApJ...555..650H}. Star formation does contribute synchrotron and free-free emission to the radio \citep{1992ARA&A..30..575C}, so caution should be noted since completely disentangling nuclear emission from circumnuclear contributions remains imprecise.

As an initial cut on the radio-loudness for the FRAMEx sample, and thereby a potential indicator for whether the radio emission may be dominated by jet activity, we have marked a radio-loudness value of $R_{\rm X}=-2.8$ in our radio-to-X-ray luminosity plots depicted in Figure \ref{fig:lrlx}. This value was determined as a potential boundary between radio-loud and radio-quiet AGNs by analyzing the statistical properties of the distributions of $R_{\rm O}$ and $R_{\rm X}$, and is more clearly defined by nuclear activity than the optical band measurements \citep{2007A&A...467..519P}. At both VLA and VLBA radio luminosities, only NGC~1052 is considered radio-loud. Indeed, NGC~1052 is the only galaxy in the sample whose host is elliptical and contains a large-scale jet that is also resolved at subparsec spatial scales \citep{1984ApJ...284..531W,2003A&A...401..113V,2024ApJ...961..109S}. The remaining hosts in the FRAMEx sample are all spiral Seyfert galaxies and do not have traditionally collimated structures  in our high-resolution VLBA observations depicted in \citetalias{2024ApJ...961..109S}, although it should be noted that NGC~4151 has an extended one-sided tail. Likewise, 17 out of 25 FRAMEx targets in our VLA imaging results depict pointlike morphology, with only 8 sources displaying extended morphology.\footnote{We consider NGC~1068, NGC~2110, NGC~2782, NGC~2992, NGC~3227, NGC~4151, NGC~4388, and NGC~5507 to have extended radio morphology.} Of FRAMEx targets with extended morphologies, only NGC~2110, NGC~4151 and perhaps NGC~4388 show linearly projected structures, but none of these terminate in larger radio lobes.

Historically, $R_{\rm O}$ has been assumed to be related to accretion disk activity which may be misleading if jet activity dominates the optical band over the accretion activity. A modern alternative classification scheme for AGNs has been proposed by \cite{2017NatAs...1E.194P}, who suggest to discriminate between ``jetted'' and ``non-jetted'' sources rather than using a radio-loud threshold. Jet-specific indicators which include direct evidence (as in NGC~1052), detection of $\gamma$-ray emission $\gtrsim 1~\rm{MeV}$, or significant excess of radio emission from the radio-infrared correlation can then be used to classify the AGN sources. In general, prominent $\gamma$-ray emission does not appear to be associated with radio-quiet AGNs, though emission components have been observed to be possibly coincident with a small number of AGNs, including the FRAMEx targets NGC~1068 and NGC~6814 \citep[][]{2012ApJ...747..104A,2015ApJ...810...14A}. In the case of NGC~6814, one of our faintest radio targets, it remains possible that the detection is associated with a background object (or is a false positive). We searched for VLBI detections in the Radio Fundamental Catalog \citep{2025ApJS..276...38P} near NGC~6814 and found only one source (J1939-1002) within a distance of five times the semimajor axis of the $\gamma$-ray detection \citep[J1942.5–1024; $0.187\degr\times0.165\degr$;][projected separation from the detection: $0.73\degr$]{2012ApJS..199...31N}. Neither AGN, however, show significant radio-infrared excess when compared to other starburst galaxies, and \cite{2012ApJ...747..104A} suggest that there is not much room for jet activity in NGC~6814. The $\gamma$-ray emission in NGC~1068 is meanwhile suggested to be attributed to its starburst component \citep{2012ApJ...755..164A}.

\subsubsection{Weak Jets and Overpressured Cocoons}
A possible explanation for the extended morphology in the radio-quiet AGNs at VLA resolution is that it is jet-induced but the jet is radiating intermittently, is weakly emitting (or young), or was not successfully launched. Jet intermittency is expected in successful launches where shells of material are launched at inconsistent velocities and subsequently collide with each other to produce internal shocks resulting in radio emission \citep{1978MNRAS.184P..61R,1999AN....320..232G}. However, if a jet expels material at launch velocities smaller than the escape velocity of the central engine, it may fall back towards the launch site and collide with newly formed material, again producing shocked emission \citep{2004A&A...413..535G}. It is possible that the ``candle flame'' shapes observed in IC~2461, NGC~3227, NGC~3786, or NGC~5899 for instance \citepalias[see Figures 1 and 2 in][]{2025ApJ...986..194S} may be due to a newly formed or failing jet, where collisions nearby the AGN are producing the one-sided whispy structure.

The progenitor of the radio emission is less clear in the several FRAMEx sources that do not show a traditional jet-like structure but have a uniquely extended morphology. For example, NGC~2782 appears to have two distinct axisymmetric morphologies surrounding the nucleus, the AGN in NGC~2992 is encapsulated within a ``figure eight''-like structure, and both NGC~3227 and NGC~5506 have radio emission that encircles the bright center. One possible explanation is that the extended structure is from the tip of a jet that interacts with and shocks the ambient medium due to a pressure imbalance \citep{1974MNRAS.166..513S,1989ApJ...345L..21B,1992ApJ...392..458C}. For host galaxies that harbor a jet, the jet beam may excavate a propagation tunnel through the nearby gas and drive a lateral pressure to form a radio ``cocoon''.

\cite{1989ApJ...345L..21B} applied observed quantities of Cygnus~A to their models of overpressured cocoons successfully, but unlike our radio-quiet targets, Cygnus~A contains a narrow jet \citep[$\sim5^{\circ}$ opening angle;][]{1984ApJ...285L..35P} whereas NGC~2782, NGC~2992, NGC~3227, and NGC~5506 contain no obvious collimated structure. While it is possible that jet filaments are faint or resolved away at VLA resolution, Cygnus~A is at a much greater distance than any of our objects, yet the jets are still apparent \citep[$z=0.057$, whereas the mean redshift of FRAMEx is $\langle z\rangle=0.007$;][]{2022ApJS..261....2K}. Furthermore, our polarimetric work on NGC~4388 \citepalias{2024ApJ...961..230S} showed that only the nuclear region was significantly polarized, albeit weakly, with the linearly polarized continuum approximately 6\% of the Stokes I continuum. This is in contrast to clearly jetted sources such as M~87 which has jet polarization along the filaments in the $20-80$\% range \citep[centered at 11 GHz;][]{2021ApJ...923L...5P}. The lack of polarized filaments in NGC~4388 is perhaps suggestive that the linear structure is not jet-like but rather a byproduct of winds interacting with the host medium, as we hypothesized.

\subsubsection{\emph{Fermi} Bubbles}
A potential solution to the lack of a defined collimated structure, however, is that the FRAMEx targets may be exhibiting radio morphology similar to that of the bipolar symmetric radio bubbles observed around the Galactic Center \citep[extending 140 parsecs $\times$ 430 parsecs;][]{2013Natur.493...66C,2019Natur.573..235H}. The edge-brightened radio structure is coincident with that of the $\gamma$-ray \textit{Fermi} bubbles at the Galactic Center, which suggests a powerful origin such as an energetic accretion event (e.g., a tidal disruption event), nuclear starburst, or past jet activity \citep{2010ApJ...724.1044S,2025arXiv250118713S}. A former jet may account for the extended structure and the lack of collimated radio emission in the FRAMEx targets, in which case the radio emission is in fact a relic of previous jet activity.

If past jet activity formed the radio emission aligned with the Galactic Center, one hypothesis is that the larger \textit{Fermi} bubbles were formed due to the precession of a former jet, perhaps due to a torqued magnetic field \citep{2010ApJ...724.1044S}. Jet precession has been observed with accreting sources on smaller scales, such as the microquasar SS~433, an X-ray bright object within the Galaxy that has a 164 day precession period along the jet axis \citep{1984ARA&A..22..507M,2004ASPRv..12....1F}. A former precessing jet in the FRAMEx case could have excavated gas in the ISM to form curved structures or the symmetrical shapes observed in NGC~2782 and NGC~2992. Indeed, magneto-hydrodynamical simulations of precessing jets in AGNs show similar ``X''- or ``S''-shaped radio morphology to NGC~2782 and NGC~2992 \citep[][]{2023ApJ...948...25N}. Jet precession is also viable in the case of highly linear radio structures such as in NGC~4151, which shows signs of wobbling \citep{1998ApJ...496..196U}, and the \cite{2023ApJ...948...25N} simulations show that the jet-precession properties may be strongly affected by the viewing angle such that the precession is not obvious. But in the case of NGC~4151 we know that the radio knots are not relativistically separating from the AGN and that some energy loss appears to dominate at greater distances from the core \citep[jet velocity $\leq0.04c$;][]{2017MNRAS.472.3842W}.

Morphological evidence for jet precession in the case of NGC~2992 is also tantalizing as its figure eight structure is strikingly similar to the simulated results. However, the figure eight radio emission structure in NGC~2992 extends roughly $8''$, or $\sim1.3$ kpc \citep[][]{1984ApJ...285..439U}, well below the resolution of the simulated results, which simulated precessing relativistic jets launching from 12 kpc cylinders.  Eventually they extend out to $\sim100$ kiloparsec scales after 96 Myr, highlighting the extreme distances at which jets can reach. Simulations with physical resolution at parsec scales have been successful at replicating radio morphology with jet precession \citep{2023ARep...67.1275T}, but again, this is a mismatch in resolution with the observed VLA morphology of NGC~2992 as its structure extends to roughly kiloparsec spatial scales. In addition to all of this, the radio loops of NGC~2992 are expected to be due to an expanding bubble and not a collimated jet \citep{2001A&A...378..787G,2005ApJ...628..113C}.

For the remainder of the sources which are pointlike and below the resolution limit of the VLA, we can only infer the radio emitting source based off of theoretical principles. It is possible that the unresolved emission may be due to a jet that is oriented towards the observer \cite[as in ][]{1979ApJ...232...34B}. Confirming jets in these sources may be tied to whether the sources have significantly polarized continuum, and magneto-hydrodynamical simulations with a high-resolution in between \cite{2023ARep...67.1275T} and \cite{2023ApJ...948...25N} may provide the additional constraints needed to determine whether jet-related activity is responsible for the radio emission. From a radio morphological perspective alone, evidence that the extended structure in the FRAMEx sample at VLA resolution is dominated by jets is middling at best and perhaps further work analyzing any polarimetric structure will help to further clarify jet-related activity.

\begin{deluxetable*}{rCCCCCCCC}
\tablecaption{AGN Wind Diagnostics}
\label{tab:winds}
\tablehead{
\colhead{Target} &
\colhead{$B_{\rm eq,min}$} &
\colhead{$B_{\rm eq,max}$} &
\colhead{${\rm KE}_{\rm min}$} &
\colhead{${\rm KE}_{\rm max}$} &
\colhead{${\rm Rad.}_{\rm min}$} &
\colhead{${\rm Rad.}_{\rm max}$} &
\colhead{$\frac{{\rm KE}_{\rm max}}{{\rm Rad.}_{\rm min}}$} &
\colhead{$L_{\rm wind}/L_{\rm Bol.}$}
\\
&
\colhead{(${\rm mG}$)} &
\colhead{(${\rm mG}$)} &
\colhead{($10^{52}~\rm erg$)} &
\colhead{($10^{52}~\rm erg$)} &
\colhead{($10^{52}~\rm erg$)} &
\colhead{($10^{52}~\rm erg$)} &
\colhead{($\%$)} &
\colhead{($\%$)} 
}
\startdata
\multicolumn{9}{c}{VLBA Detection} \\
\hline
NGC~1052 &    0.13 & 15 & 2.7\times10^{-3}  &  3.3               & 1.2  & 1439   &  274    &    257 ^{}_{} \\
NGC~1068 &     1.6 & 16 & 4.7\times10^{-5}  &  1.5\times10^{-3}   & 0.59 & 19     &    0.26 &    9.2 ^{}_{} \\ 
NGC~2110 &    0.54 & 49 & 3.5\times10^{-5}  &  3.0\times10^{-2}  & 5.1  & 4446   &    0.59 &   0.78 ^{+0.17}_{-0.16} \\
NGC~2782 &    0.11 & 19 & 2.7\times10^{-5}  &  6.2\times10^{-2}  & 0.80 & 1852   &    7.8  &   3.8  ^{+8.3 }_{-2.3 } \\
NGC~2992 &    0.15 & 16 & 1.1\times10^{-4}  &  0.12              & 2.9  & 3089   &    4.02 &   4.2  ^{+0.3 }_{-0.3 } \\
NGC~3079 &   0.076 & 10 & 3.6\times10^{-4}  &  0.53              & 0.28 & 408    &  192    &    148 ^{}_{} \\
NGC~4151 &    0.87 & 86 & 6.5\times10^{-6}  &  6.4\times10^{-3}  & 0.70 & 677    &    0.91 &   1.1  ^{+0.2 }_{-0.2 }\\
NGC~4235 &    0.22 & 15 & 2.2\times10^{-5}  &  1.2\times10^{-2}  & 3.9 & 2201    &    0.31 &   0.63 ^{+0.17}_{-0.17} \\
NGC~4388 &    0.53 & 44 & 2.0\times10^{-6}  &  1.6\times10^{-3}  & 1.3 & 954     &    0.12 &   0.19 ^{+0.11}_{-0.04} \\
NGC~4593 &    0.38 & 25 & 4.3\times10^{-6}  &  2.3\times10^{-3}  & 4.6 & 2415    &    0.05 &   0.11 ^{+0.02}_{-0.02} \\
NGC~5290 &    0.14 & 15 & 3.1\times10^{-5}  &  3.5\times10^{-2}  & 1.6 & 1758    &    2.2  &   2.3  ^{+0.5 }_{-0.4}\\
NGC~5506 &    0.37 & 40 & 1.3\times10^{-4}  &  0.15              & 2.7 & 3059    &    5.5  &   5.4  ^{+1.5 }_{-0.7}\\
\hline
\multicolumn{9}{c}{VLBA Detection} \\
\hline
NGC~1320 &    0.12 & \nodata & \nodata           &  3.7\times10^{-2}  & \nodata  & 2361    & \nodata &   1.8  ^{}_{} \\
 IC~2461 &    0.16 & \nodata & \nodata           &  6.6\times10^{-3}  & \nodata  & 1452    & \nodata &   0.51 ^{}_{} \\
NGC~3081 &    0.13 & \nodata & \nodata           &  1.8\times10^{-2}  & \nodata  & 3325    & \nodata &   0.62 ^{+0.17}_{-0.23} \\
NGC~3089 &    0.25 & \nodata & \nodata           &  2.8\times10^{-3}  & \nodata  & 4440    & \nodata &   0.07 ^{+0.04}_{-0.02} \\
NGC~3227 &    0.34 & \nodata & \nodata           &  1.4\times10^{-2}  & \nodata  & 707     & \nodata &   2.3  ^{+0.2 }_{-0.2}\\
NGC~3786 &    0.19 & \nodata & \nodata           &  1.6\times10^{-2}  & \nodata  & 1217    & \nodata &   1.5  ^{}_{}\\
NGC~4180 &    0.13 & \nodata & \nodata           &  9.6\times10^{-3}  & \nodata  & 1387    & \nodata &   0.79 ^{}_{} \\
NGC~5899 &    0.16 & \nodata & \nodata           &  5.2\times10^{-2}  & \nodata  & 2109    & \nodata &   2.8  ^{+0.7 }_{-0.6 }\\
NGC~6814 &    0.18 & \nodata & \nodata           &  5.4\times10^{-3}  & \nodata  & 1688    & \nodata &   0.37 ^{+0.04}_{-0.04} \\
NGC~7314 &    0.18 & \nodata & \nodata           &  7.7\times10^{-3}  & \nodata  & 1249    & \nodata &   0.70 ^{+0.06}_{-0.06} \\
NGC~7378 &    0.15 & \nodata & \nodata           &  2.3\times10^{-3}  & \nodata  & 1597    & \nodata &   0.17 ^{+1.92}_{-0.07} \\
NGC~7465 &    0.15 & \nodata & \nodata           &  1.3\times10^{-2}  & \nodata  & 1312    & \nodata &   1.2  ^{+0.2 }_{-0.1}\\
NGC~7479 &    0.16 & \nodata & \nodata           &  3.4\times10^{-2}  & \nodata  & 1534    & \nodata &   2.5  ^{}_{}
\enddata
\tablecomments{Column (1) target name. Column (2) Lower limit on magnetic field strength using VLA point source, assuming equipartition from VLA excess emission. Column (3) Upper limit on magnetic field strength using VLBA point source, assuming equipartition from VLA excess emission. Column (4) Minimum mechanical energy converted into synchrotron radiation (from excess emission in Table \ref{tab:diagnostics}) using $B_{\rm eq,max}$. Column (5) Maximum mechanical energy converted into synchrotron radiation (from excess emission in Table \ref{tab:diagnostics}) using $B_{\rm eq,min}$. Column (6) Minimum radiative output of nucleus, assuming 10\% of the bolometric luminosity is converted into AGN winds over the synchrotron lifetime with a magnetic field strength of $B_{\rm eq,max}$. Column (7) Maximum radiative output of nucleus, assuming 10\% of the bolometric luminosity is converted into AGN winds over the synchrotron lifetime with a magnetic field strength of $B_{\rm eq,min}$. Column (8) Maximum kinetic energy as a percentage of the minimum radiative wind output. Column (9) Estimated wind luminosity from VLA excess emission as a percentage of bolometric luminosity. Wind luminosity is estimated as $L_{\rm wind}=L_{\rm excess}/\eta_{\rm wind}$ where $\eta_{\rm wind}=3.6\times10^{-5}$ \citep{2014MNRAS.442..784Z}. }
\end{deluxetable*}

\subsection{Outflows \& Winds}

The well-known tight correlation between the mass of the central SMBH and the velocity dispersion of the surrounding stellar bulge \citep[$M\propto\sigma^{\alpha}$;][]{2000ApJ...539L...9F,2000ApJ...539L..13G} suggests an intricate connection between the central engine and the host galaxy. This is despite the broad extent of the enveloping host bulge \citep[typically $<500~{\rm pc}$;][]{2002ApJ...578...90F} relative to the essentially negligible gravitational sphere of influence of the SMBH, which is on the order of tens of parsecs and presumably affects the motion of only nearby stars \citep[][]{2004cbhg.symp..263M,2015ARA&A..53..115K}. The implication is that the SMBH co-evolves with the bulge such that the accretion process must transfer some of its energy to its immediate surroundings \citep{1999MNRAS.308L..39F}. Quasi-spherical outflows in the form of intense winds have been suggested as a potential source of energy transfer from the accretion disk, and these may sweep up nearby gas as they expand outward from the central source \citep{1998A&A...331L...1S,2003ApJ...596L..27K,2012ApJ...753...75C}.

High-velocity ionized winds appear to be relatively common and have speeds reaching several thousand kilometers per second \citep{2015ARA&A..53..115K,2014MNRAS.442..784Z}. These uncollimated outflows exert force upon discontinuities in the ISM, shocking the ambient medium to produce synchrotron radiation observed in the radio \citep[][see also Figure 1 in \citealt{2015MNRAS.447.3612N}]{2014MNRAS.442..784Z,2019ApJ...887..200F,2023ApJ...953...87F,2020ApJ...904...83S,2019NatAs...3..387P,2024Galax..12...17H}. The AGN-driven winds function similarly to winds injected from supernova events \citep{2012MNRAS.425..605F,1988ApJ...334..252C}, but \cite{2017A&A...601A.143F} find that winds from populations of supernovae do not significantly contribute to the observed AGN wind luminosity.

The bolometric luminosity of an AGN can be approximated to have $0.1-10\%$ of the AGN power converted into accretion disk wind luminosity \citep{2017A&A...601A.143F,2013MNRAS.436.2576L,2014MNRAS.442..784Z}. Using these conversions, we suggested in \citetalias{2024ApJ...961..230S} that the nuclear energy budget in NGC~4388 could sufficiently supply FRAMEx AGN winds observed as synchrotron emission if it was produced from wind interactions shocking the host ISM, and we hypothesized that winds may be responsible for a complex ionization cone directly to the south of the AGN. The ionized region is coincident with measurements of the shock emission indicator [\ion{Fe}{2}] \citep{2001AJ....122..764K,2017MNRAS.465..906R}. The wind interpretation is supported by the relative edge-on nature of NGC~4388, where measurements of [\ion{Fe}{2}] and [\ion{O}{3}] in the optical are reduced outside of the cone north of the AGN, perhaps due to the host disk obscuring in the line of sight. Furthermore, enhanced [\ion{Fe}{2}] emission is observed to be spatially coincident with a secondary bright radio knot that, when combined with its steep spectrum ($\alpha=-1.07\pm0.11$), provides additional support for strong AGN winds interacting with the host medium. The combination of all of these factors led us to suggest that the ionization region may in fact be emission projected onto the underside of the host disk rather than a typical jet interpretation.

\subsubsection{Energy Budget}
In our \citetalias{2024ApJ...961..230S} estimates for NGC~4388, we compared nuclear wind energetics to circumnuclear mechanical feedback, assuming a shocked ISM in the host environment which produces synchrotron radiation. The mechanical energy was measured using ${\rm KE}=2N_{\rm e}m_{\rm e}c^2(\gamma-1)$. Here, $N_{\rm e}$ is the number of electrons and was determined by dividing the observed luminosity by the mean total power of the synchrotron emission, $\langle P\rangle=\frac{4}{3}\sigma_{\rm T}\beta^2\gamma^2cU_B$, with $\gamma$ being the Lorentz factor and $U_B=\frac{B^2}{8\pi}$ being the magnetic energy density. We estimated $\gamma\approx\sqrt{\frac{2\pi m_{\rm e}c\nu}{eB}}$, where $\nu$ is the observing frequency. To determine the electron count we use the 4.4 GHz luminosities from the excess flux densities listed in Table \ref{tab:diagnostics} and assume that all of the energy measured from the nuclear excess emission observed by the VLA is is attributed to wind interactions. With these calculations we have extended the analysis of wind energetics to the full sample contained entirely within the resolving limits of the VLA (i.e. the wind energetics associated with the nuclei), albeit with some caveats on the magnetic field strengths used.

For the purpose of approximating the nuclear magnetic field strengths, we use $B_{\rm eq}=0.27R^{-1}_{\rm pc}L^{1/2}_{46}~{\rm G}$ \citep[equation 21 in ][]{2008MNRAS.390..847L}. The equation for $B_{\rm eq}$ is derived from an isotropically emitting synchrotron source with a uniform magnetic field strength and a uniform distribution of relativistic electrons. $R_{\rm pc}$ is the size of the radio-emitting sphere in units of parsecs and $L_{46}=L_{\rm Bol.}/10^{46}$. The equation assumes equipartition from a shared energy source between the magnetic fields and the associated cosmic-ray electrons, which in this case may be a shocked ISM. Simulations for star-forming galaxies by \cite{2019Galax...7...45S} suggest that the equipartition assumption is likely not valid on length scales smaller than the mean propagation length of the cosmic-ray electrons which is determined by turbulence in the ISM. For star-forming spiral galaxies this turbulence is primarily driven by supernovae and extends to length scales approximately $50-100$ pc in size, similar to the physical scales probed by the VLA. \cite{2019Galax...7...45S} further note that deviations from equipartition may occur in a highly magnetized environment where severe energy losses may be more prominent for cosmic-ray electrons than cosmic-ray protons, such as active or star-bursting nuclei. Equipartition is however predicted in AGNs when gas clouds are shocked by a nearby synchrotron self-absorbed compact jet \citep{1979ApJ...232...34B}, the modeling of which is supported by observations \citep{2011A&A...532A..38S}. For our purposes, we assume that this also extends to accretion disk winds shocking a turbulent ISM. We constrain the size of the radio emitting source $R_{\rm pc}$ with limits between the projected diameters of the synthesized beams from the VLA and VLBA imaging results (listed as $d_{\rm VLA}$ and $d_{\rm VLBA}$ in Table \ref{tab:diagnostics}), assuming that all of the AGN wind energetics are produced within this range. Using these constraints on $R_{\rm pc}$, we indicate both an upper and lower limit of the magnetic field strength and the kinetic energy for all targets in Table \ref{tab:winds}.

\begin{figure}[ht!]
    \centering
    \subfigure[FRAMEx Wind Energetics][Wind energetics]{
        \label{fig:winds}\includegraphics[width=\columnwidth]{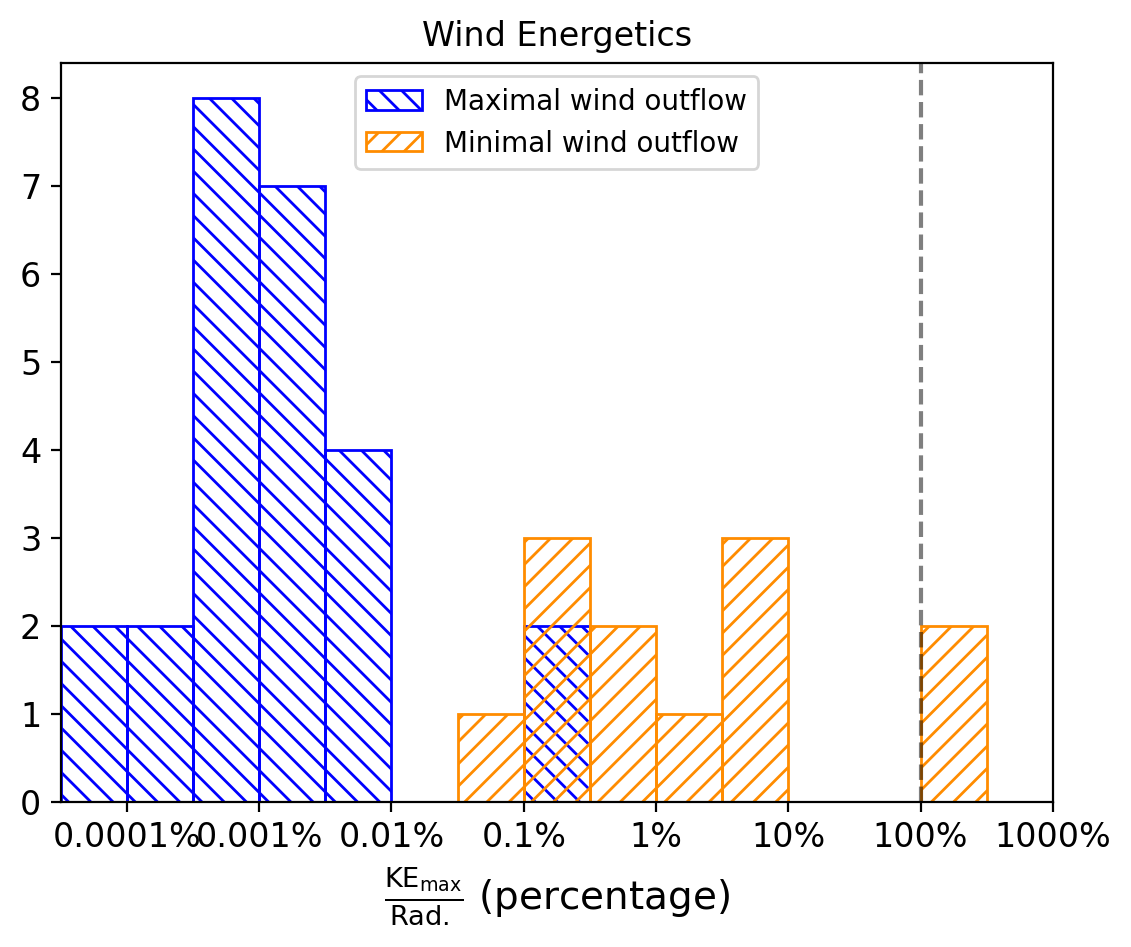}
    } \\
    \subfigure[SFRs][Excess radio SFR versus narrow-line H$\alpha$ SFR (maximum)]{
        \label{fig:sfrs}\includegraphics[width=\columnwidth]{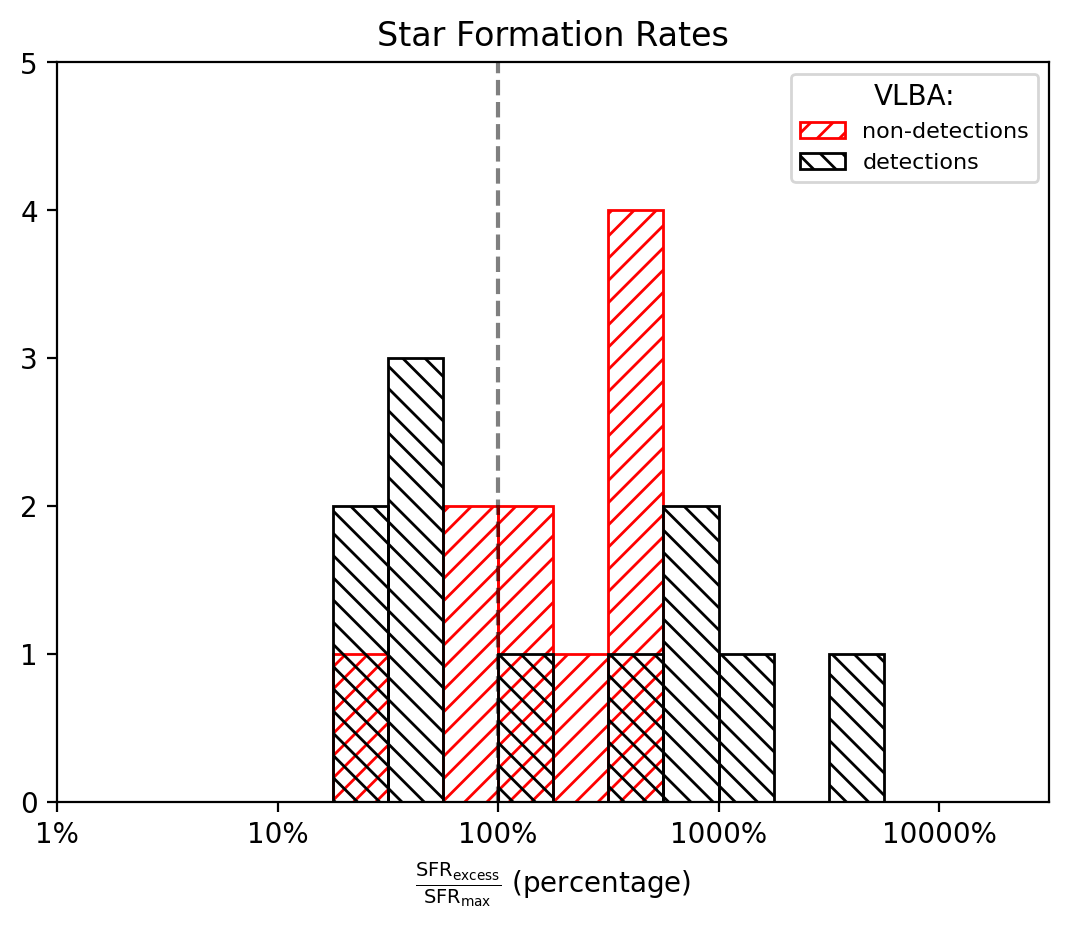}
    }
    \caption{\subref{fig:winds} Distribution of wind energetics. Orange indicates the circumnuclear radio energies attributed to wind interactions as a percentage of the minimal wind outflow ($\frac{\rm KE_{max}}{\rm Rad._{min}}$). The blue indicates the circumnuclear radio energies attributed to wind interactions as a percentage of the maximal wind outflow ($\frac{\rm KE_{max}}{\rm Rad._{max}}$). \subref{fig:sfrs} Excess radio SFRs versus narrow-line H$\alpha$ SFRs. Red are the VLBA non-detected FRAMEx sources and black are the VLBA detected sources.}
    \label{fig:windssfrs}
\end{figure}

As in \citetalias{2024ApJ...961..230S}, we estimate the total energy of the expected wind luminosity by multiplying it by the synchrotron lifetime of cosmic ray electrons,
\begin{equation}
t_{\rm synch}\approx1.06\times10^9~{\rm yr}\left(\frac{B}{\rm \mu G}\right)^{-\frac{3}{2}}\left(\frac{\nu}{\rm GHz}\right)^{-\frac{1}{2}}.
\end{equation}
The range of energetics of the radiative output assuming a wind efficiency of 10\% of that of the bolometric luminosity, i.e. $0.1L_{\rm Bol.}t_{\rm synch}$, are indicated for all FRAMEx targets in Table \ref{tab:winds}. In Figure \ref{fig:winds} we depict distributions as percentages of the excess radio energetics attributed to both a minimal and maximal wind outflow. We find that the synchrotron energy exceeds the radiative output only in the cases of NGC~1052 and NGC~3079, the two most radio-luminous objects in the sample\footnote{NGC~1052 is a radio-loud AGN with large-scale jet structure that extends 2.8 kpc \citep{1984ApJ...284..531W,2020AJ....159...14N}, while NGC~3079 has been proposed to be a radio-loud AGN as well \citepalias{2023ApJ...958...61F}.}. But in these two cases the kinetic response only barely surpasses the minimum radiative output due to winds and is several orders of magnitude smaller than the maximum. The remaining targets have radiative output that can sufficiently supply the mechanical energy observed in all cases if it is indeed sourced from wind-induced synchrotron emission, potentially indicating that AGN winds may be a dominating source of the radio luminosity.

\subsubsection{Calculating Wind Luminosity from Radio Emission}
Finally, \cite{2014MNRAS.442..784Z} proposed that AGN winds perhaps undergo similar energetic conversions to that of starburst-driven winds in a normal star-forming galaxy. Using results from galactic evolution models, they suggested an efficiency factor for the 1.4~GHz radio luminosity that scales with the wind luminosity as $\eta_{\rm wind}=3.6\times 10^{-5}$, with $L_{\rm wind}=L_{\rm 1.4~GHz}/\eta_{\rm wind}$. They calculated a median value of the wind luminosity for radio quiet quasars as 4\% of the bolometric luminosity, within agreement of the expected range of $0.1-10\%$. We calculated the wind luminosities from this conversion using our 4.4~GHz radio luminosities (determined from the excess flux densities listed in Table \ref{tab:diagnostics}) assuming a relatively flat spectrum down to 1.4~GHz, and we list the ratios of $L_{\rm wind}/L_{\rm Bol.}$ as percentages in Table \ref{tab:winds}. The two most radio-luminous sources in the FRAMEx sample have wind luminosities that overestimate the bolometric output, NGC~1052 (257\%) and NGC~3079 (148\%), closely mimicing the energy budget calculation. This perhaps indicates that radio emission in these sources is a combination of other processes such as star formation (see next section) or coronal emission. The remaining targets have wind luminosities ranging between $0.07-5.4\%$, in line with the expected range of $0.1-10\%$ of $L_{\rm Bol.}$.

\begin{deluxetable*}{rCCCCC}
\tablecaption{Star Formation Diagnostics}
\label{tab:SFRs}
\tablehead{
\colhead{Target} &
\colhead{$\rm FWHM_{ H\alpha}$} &
\colhead{${\rm SFR_{VLA}}$} &
\colhead{${\rm SFR_{excess}}$} &
\colhead{$\rm{SFR_{max}}$} &
\colhead{$\frac{\rm SFR_{excess}}{\rm SFR_{max}}$}
\\
&
\colhead{(${\rm km~s^{-1}}$)} &
\colhead{(${M_{\odot}~{\rm yr}^{-1}}$)} &
\colhead{(${M_{\odot}~{\rm yr}^{-1}}$)} & 
\colhead{(${M_{\odot}~{\rm yr}^{-1}}$)} & 
\colhead{}
}
\startdata
\multicolumn{6}{c}{VLBA Detection} \\
\hline
NGC~1052 & 633 \pm 1  &    36 \pm 2    & 40    \pm 7    &    0.11  &    365 \\
NGC~1068 & 995 \pm 1  &  0.36 \pm 0.05^{a} &  0.77 \pm 0.02 &    1.7   &      0.44\\
NGC~2110 & 674  \pm 1 &   3.5 \pm 0.2  &  3.0  \pm 0.6  &    0.38  &      8.0 \\
NGC~2782 & \nodata    &  0.72 \pm 0.04 &  0.58 \pm 0.04 &    2.6   &      0.22 \\
NGC~2992 & 420  \pm 4 &   1.4 \pm 0.1  &  1.7  \pm 0.1  &    0.10  &     17 \\
NGC~3079 & 769  \pm20 &  3.9  \pm 0.2  &  2.9  \pm 0.4  &    0.09  &     32 \\
NGC~4151 & 448  \pm 0 &  1.8  \pm 0.1  &  1.3  \pm 0.1  &    1.3   &      1.0 \\
NGC~4235 & 451  \pm 4 &  0.19 \pm 0.01 &  0.31 \pm 0.07 &    0.68  &      0.45 \\
NGC~4388 & 404  \pm 0 &  0.11 \pm 0.01 &  0.15 \pm 0.01 &    0.55  &      0.28 \\
NGC~4593 & 374  \pm11 &  0.17 \pm 0.01 &  0.14 \pm 0.03 &    0.35  &      0.39 \\
NGC~5290 & 385  \pm 3 &  0.81 \pm 0.04 &  0.49 \pm 0.08 &    0.10  &      4.9 \\
NGC~5506 & 439  \pm 1 & 19    \pm 1    &  8.3  \pm 0.9  &    1.2   &      7.0 \\
\hline
\multicolumn{6}{c}{VLBA Non-detection} \\
\hline
NGC~1320 & \nodata    &  0.36 \pm 0.03  & \nodata  &    0.24  &    1.5  \\
 IC~2461 & 336  \pm 4 &  0.09 \pm 0.01  & \nodata  &    0.02  &    4.3  \\
NGC~3081 & 356  \pm 1 &  0.17 \pm 0.02  & \nodata  &    0.74  &    0.23 \\
NGC~3089 & \nodata    &  0.09 \pm 0.01  & \nodata  & \nodata  & \nodata \\
NGC~3227 & 572  \pm 4 &  0.88 \pm 0.04  & \nodata  &    0.48  &    1.8  \\
NGC~3786 & 345  \pm 3 &  0.38 \pm 0.10  & \nodata  &    0.42  &    0.91 \\
NGC~4180 & \nodata    &  0.10 \pm 0.01  & \nodata  & \nodata  & \nodata \\
NGC~5899 & 531  \pm 1 &  1.01 \pm 0.05  & \nodata  &    1.2   &    0.82 \\
NGC~6814 & 314  \pm 6 &  0.10 \pm 0.01  & \nodata  &    0.02  &    4.9  \\
NGC~7314 & 211  \pm 1 &  0.19 \pm 0.01  & \nodata  &    0.04  &    4.8  \\
NGC~7378 & \nodata    &  0.03 \pm 0.01  & \nodata  & \nodata  & \nodata \\
NGC~7465 & \nodata    &  0.25 \pm 0.02  & \nodata  &    0.22  &    1.1  \\
NGC~7479 & 384  \pm 2 &  0.37 \pm 0.02  & \nodata  &    0.08  &    4.6 
\enddata
\tablecomments{Column (1) Target name. Column (2) Full width at half maximum for emission-line measurement of narrow H$\alpha$ spectral region \citep{2017ApJ...850...74K}. Column (3) SFR based on VLA measurement. Column (4) SFR based on extranuclear radio emission. Column (5) Maximum SFR from narrow H$\alpha$ emission-line spectral region \citepalias{2021ApJ...906...88F}. Column (6) Ratio of SFR from excess VLA emission and maximum SFR from narrow-line H$\alpha$ luminosity.\\
$^a$ We used the S1 flux density from \cite{2025MNRAS.539..808M} and a spectral index of 0 \citep{2024MNRAS.52711756M} to calculate the SFR for NGC~1068.}
\end{deluxetable*}

\subsection{Star Formation}

Recent ($<10^8~{\rm yr}$) and intense star-formation activity ionizes the interstellar medium (ISM) and produces the \ion{H}{2} regions commonly observed in nearby spiral galaxies, including regions within their cores \citep[][]{1983ApJ...268L..79T,1992ARA&A..30..575C}. In the radio, \ion{H}{2} regions are typically characterized by a spectral index of $\alpha\approx-0.7$ \citep[e.g.][]{1988A&A...195...38V,1992ARA&A..30..575C,2011ApJ...737...67M}, which, as mentioned in \citetalias{2025ApJ...986..194S}, is remarkably similar to the mean spectral index of $\langle\alpha^{\rm ND}_{\rm CX}\rangle=-0.69$ measured for the FRAMEx targets not detected by the VLBA, indicating that these regions could perhaps be dominated by star-forming processes. Diagnostics for measuring star formation rates (SFRs) have been formulated across the electromagnetic spectrum, but radio emission is particularly useful because it is largely unaffected by dust attenuation \citep{2011ApJ...737...67M,2001ApJS..133...77H}. However, expressions for SFR estimates have historically been calibrated with normal (non-active) galaxies such as NGC~6946, the late-type spiral galaxy with several star-forming regions and a starbursting nucleus \citep{2011ApJ...737...67M,2006AJ....132.2383T}. An issue with using these SFR calibrations on the FRAMEx sample are the active nuclei, where the AGN accretion process may produce emission that outshines and dilutes light from any corresponding stellar population \citep{2001ApJ...546..845G,2009ApJ...702..441G}. Thus, if one assumes contemporaneous stellar output, nuclear emission may potentially skew the interpreted SFRs toward erroneously higher than expected measurements \citep{2001ApJ...554..803Y}.

Given the similarity of the spectral indices in normal star-forming galaxies and the FRAMEx nondetections, we measure the radio-traced SFRs near the nucleus using the calibration from \citet[equation 15]{2011ApJ...737...67M} for the complete FRAMEx sample, listed in Table \ref{tab:SFRs}. We assume that all of the core VLBA emission is produced by the AGN directly and that the excess flux density measured by the VLA is entirely converted into star-forming processes. We used the difference in flux densities between our VLA and VLBA results at 4.4 GHz to determine the circumnuclear SFRs ($S_{\rm excess}$ in Table \ref{tab:diagnostics}) assuming a typical spectral index of $-0.7$ for star formation in the VLBA-detected sources. To determine whether the nuclear radio emission is too luminous for star-forming processes alone, we compare the radio-traced measurements to upper limits on nuclear SFRs traced by the reddening-corrected narrow H$\alpha$ emission-line luminosity \citepalias[][also listed in this work as $\rm{SFR}_{\rm max}$ in Table \ref{tab:SFRs}]{2021ApJ...906...88F}. When corrected for dust extinction, H$\alpha$ emission is an excellent indicator of nearby ($z\lesssim0.5$) young (stellar ages $\lesssim20~\rm{Myr}$) and massive (stellar masses $\gtrsim10~M_{\odot}$) star formation as its observation effectively probes unobscured ionization within stellar populations if one assumes that the line emission is chiefly due to the star-forming processes \citep{1998ARA&A..36..189K,2011ApJ...737...67M}.

The distribution of ratios measured for SFRs from the excess radio emission compared to the narrow-line H$\alpha$ are depicted in Figure \ref{fig:sfrs} and listed in Table \ref{tab:SFRs}. We find that 13 out of 22 FRAMEx targets overestimate the corresponding $\rm{SFR_{max}}$ measurement when using the excess radio emission marker to determine the extranuclear radio SFR ($S_{\rm excess}$ in Table \ref{tab:diagnostics}). An additional 3 sources account for more than 80\% of the maximum. The remaining targets account for less than 50\% of $\rm{SFR_{max}}$, consisting of five VLBA-detected targets (NGC~1068, NGC~2782, NGC~4235, NGC~4388, and NGC~4593) and one non-detection (NGC~3081), which may be dominated by star forming processes.

However, in the FRAMEx sample the accretion process is likely playing a significant role in the output of the H$\alpha$ luminosity, particularly in Type 1 nuclei where both narrow- and broad-line H$\alpha$ luminosity is diluting any starlight due to the lack of obscuring tori \citep{2006ApJ...641..117G,2009ApJ...702..441G}. Thus, FRAMEx objects which contain starbursting nuclei that dominate over the accretion process potentially reside with Type 2 nuclei, but only if they contain unaccounted for excess optical continuum emission \citep[][]{2001ApJ...546..845G}. The Type 2 objects which do not overestimate their respective maximum SFRs are NGC~2782 and NGC~4388 for the VLBA-detected sources and only NGC~3081 for the non-detections. 

As a point of comparison, we have included FWHM values from the narrow H$\alpha$ emission-line spectral region in Table \ref{tab:SFRs}, which range from $211-995~{\rm km~s^{-1}}$ with a mean value of $475~{\rm km~s^{-1}}$ \citep{2017ApJ...850...74K} to verify whether star formation can produce such H$\alpha$ signatures. We use these kinematics to compare to two extremes of star formation. On the low-end, we refer to small clusters of star formation in the dwarf irregular galaxy NGC~1569 \citep{2022MNRAS.513.1755S}, where winds from supernova remnants produce expansion velocities of strong H$\alpha$ in the range of $87-188~{\rm ~km~s^{-1}}$. This implies that if all outflows at the center of the AGNs were due to star formation we could expect to first order to have measured FWHMs of $\sim140~{\rm ~km~s^{-1}}$. We define the upper end of the range using the prototypical star-forming galaxy M82, which has no detectable AGN \citep[${\rm SFR}\approx10M_{\odot}~{\rm yr^{-1}}$ over the last 50 Myr;][]{1992ARA&A..30..575C,2001A&G....42d..12D}. M82 has a starbursting nucleus with H$\alpha$ line velocities of $\sim300~{\rm ~km~s^{-1}}$ that span a 1 kiloparsec region. Thus if we assume that strong star formation was occurring in the nuclear regions of FRAMEx targets described here, we would at minimum expect narrow-line H$\alpha$ kinematics to be in the range of $\sim140-300~{\rm ~km~s^{-1}}$. We find that FHWMs for the FRAMEx targets are in-line or surpass this range, but not significantly so.

\begin{figure}[ht!]
    \centering
    \includegraphics[width=\columnwidth]{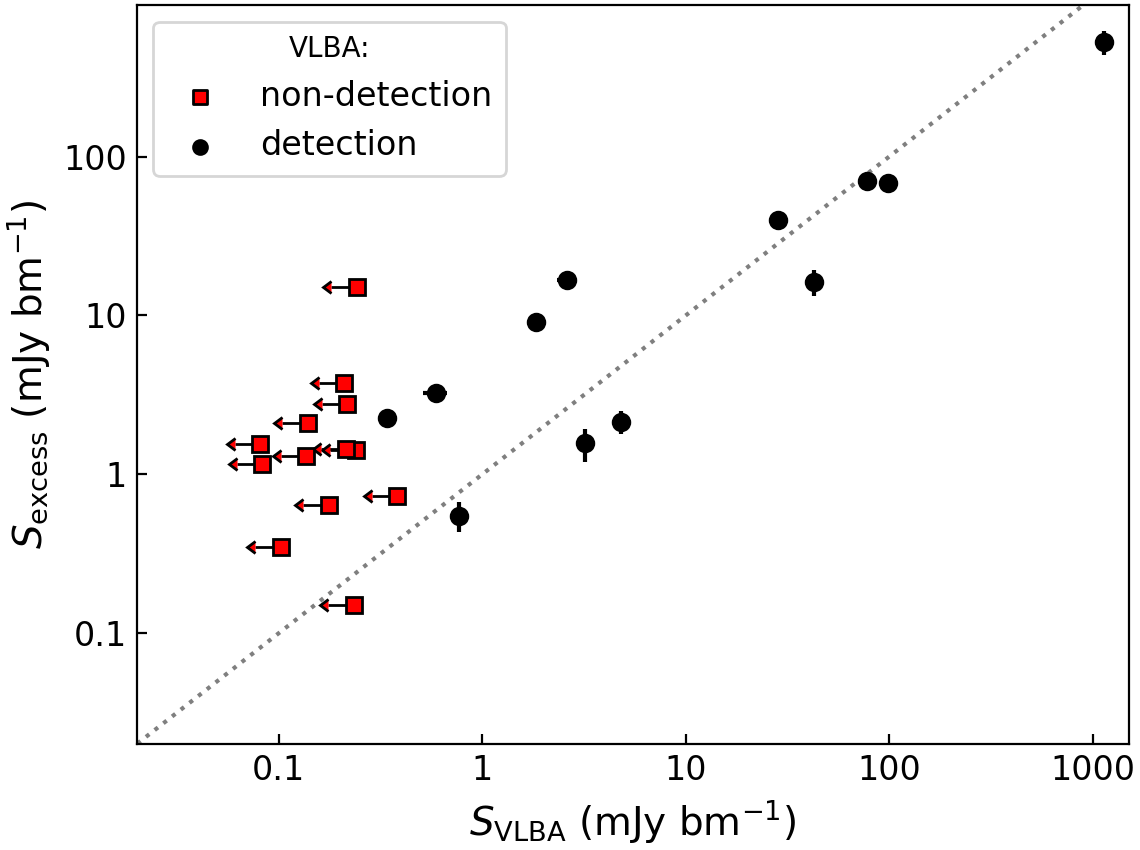}
    \caption{Excess VLA emission as a function of VLBA source brightness with upper limits indicated for VLBA non-detections (see Table \ref{tab:framex}). The dotted line indicates a one-to-one ratio.}
            \label{fig:excess}
\end{figure}

In Figure \ref{fig:excess} we plot the excess VLA flux densities against the VLBA brightness levels and find that the excess emission correlates quite well with the VLBA brightness. 
This correlation combined with the SFR overestimation may imply that the nuclear region is not dominated by star formation in the FRAMEx sample. On the other hand, the H$\alpha$ line widths support star formation as plausibly contributing to the nuclear narrow H$\alpha$ emission-line. Overall, this suggests that the FRAMEx targets have radio emission that may be comprised of several emission production mechanisms.

\section{Summary and Conclusions} \label{sec:conclusion}

This paper provides detailed diagnostics of the radio emission mechanisms of all AGNs between the declinations of $-30^{\circ}$ and $60^{\circ}$ within a 40 Mpc volume. We have previously observed the entire sample of 25 AGNs using the VLBA to probe the subparsec spatial scales but have thus far detected only 12 out of 25 objects. This is despite the fact that detections were expected based on the fundamental plane of black hole activity, which relates the X-ray and radio luminosities with the black hole mass \citep{2003MNRAS.345.1057M}. In this work, we have analyzed multiple radio emission mechanisms as they relate to the entire FRAMEx sample using a uniform set of VLA observations. These observations have a 100\% detection rate and probe regions spanning $\sim30-110~{\rm pc}$ in projected sizes, much larger than our VLBA observations which probe $\sim0.3-1~{\rm pc}$. We use these VLA observations to gauge why a majority of the targets remain undetected at the subparsec spatial scales when probed by the VLBA.

We aim to constrain the source of the radio emission properties observed by the VLA but not the VLBA by analyzing the characteristics of the high and low resolution radio imaging results, the source morphology, and how the radio emission relates to X-ray attributes. Our main conclusions as it pertains to radio emission mechanisms at VLA resolution are as follows.

\subparagraph*{Coronal activity.} The entire FRAMEx sample has theoretical size scales of the radio emitting source that are compatible with coronal origins, which can extend to tens of light days. However, the objects not detected by the VLBA have VLA-scale emission that is likely produced beyond parsec spatial scales and thus is unlikely to be due to radio emission with coronal origins. Sources detected with the VLBA are compatible with coronal emission based on their consistency with the Güdel-Benz relation ($L_{\rm R}/L_{\rm X}\sim10^{-5}$), assuming it also accurately portrays AGN coronal activity. The exception to this may be NGC~1052, which has $L_{\rm R}/L_{\rm X}\sim10^{-2.1}$, several orders of magnitude higher than the Güdel-Benz relation.  Based on their spectra, only NGC~1052, NGC~2110, and NGC~4235 have inverted spectra which may be consistent with coronal emission at $\sim\rm GHz$ frequencies according to simulations by \cite{2016MNRAS.459.2082R}. Two targets (NGC~2782 and NGC~5506) have steep spectra that are not consistent with coronal emission, while the remainder of VLBA detected targets have flat or negative spectra and the contributions from coronal origins remain inconclusive. We also measured the hard X-ray variability and find that sources detected by the VLBA are marginally more likely to have statistically significant variability with a $p$-value of 0.05.

\subparagraph*{Jets or bubbles.} NGC~1052 is the only target with a clearly defined large-scale jet. It also happens to be the brightest source of the sample by orders of magnitude and is the only source that has an elliptical host. All the other FRAMEx targets are within disk-based hosts. NGC~2110 and NGC~4151 have a linearly projected morphology which may be similar to the \textit{Fermi} bubbles at the center of our Milky Way galaxy. Other FRAMEx targets with morphology have structures that are not consistent with jets, but it is possible that jet precession, a young jet, or failed jet processes are responsible for the radio structures.

\subparagraph*{AGN winds.} We estimate radiative output energetics  if 10\% of the bolometric luminosity is converted into AGN winds, and compare these to the energetics measured from the radio synchrotron output. This calculation assumes that the excess emission observed by the VLA but not by the VLBA is all due to wind activity. Only the two brightest sources in the sample, NGC~1052 and NGC~3079, have maximum kinetic energies measured with radio emission that (just barely) exceeds the minimum radiative output if it is entirely converted into AGN winds. This suggests that all targets have excess radio emission properties that are compatible with AGN winds.

\subparagraph*{Star formation.} We measure SFRs based on the excess radio emission for the full sample and compare it to upper limits on star formation from the narrow-line H$\alpha$ emission. In a majority of the cases, the radio SFRs overestimate the maximum expected SFRs from the H$\alpha$ optical tracer. The H$\alpha$ kinematics are however in line with star formation expectations from highly star-forming galaxies that do not contain AGNs. This implies that the excess radio emission is too luminous for star formation alone, but star formation may contribute to the VLA-scale nuclear emission in smaller capacities.
\newline

The analyses presented here assume that all of the radio emission observed by the VLA is due to a singular source. While it is possible that the radio emission observed with the VLA, or the excess emission between the VLA and VLBA scales, is dominated by one single process, fully interpreting its origins remains complex. Multi-wavelength studies are crucial for disentangling the potential sources of the radio emission, particularly for sources with extended radio morphology.

Spatially resolved maps of ionized gases near AGNs in a volume-complete sample such as FRAMEx may help to decipher whether the radio emission is largely a byproduct of a shocked host medium or due to a pure synchrotron source such as a jet. Examples include NGC~4388, where the extended radio morphology is spatially coincident with ionized structure south of the nucleus, particularly with the shock indicator [\ion{Fe}{2}], and \citetalias{2024ApJ...961..230S} suggests that the extended structure is the culmination of feedback due to AGN winds shocking the host medium. Another case is NGC~1068, in which \cite{2023ApJ...953...87F} found the radio emission surrounding the nucleus to be intertwined along the edge of the narrow-line region (the region where AGN gas is ionized), where molecular gas reservoirs were rotating into the AGNs ionizing radiation field. The observed morphology was thus suggested to be a byproduct, likely formed by an interaction from powerful winds at small radii with a dense surrounding medium. Careful analysis with a volume-complete survey comparing radio extents in radio-quiet AGN and the associated outflowing or disturbed ionized gas kinematics would confirm or refute whether the radio emission is produced in situ.

There is a reserve of unresolved structure in the radio between $\sim1-100~{\rm pc}$ that neither the VLBA nor the VLA are able to observe (between the VLA's longest baseline in A-config compared to the VLBA's shortest baseline). Exploring the entire FRAMEx sample at these intermediate spatial scales with detailed morphological studies that can fill the missing spatial scales will help to further constrain the source of the radio emission by showcasing any extended structure that is resolved away when observed with the VLBA but unresolved with the VLA. For example, jet interpretations of the radio emission may be supported if higher-resolution imaging depicts linearly projected features. On the other hand, AGN wind production may be supported if no new morphology is detected at the intermediate spatial scales.

Such observations may include using the \textit{e}-MERLIN interferometer to further close the gap, where a 5 GHz observation with $0\farcs05$ resolution can probe down to $\sim10$ pc spatial scales at the 40 Mpc limiting distance of FRAMEx. A recent example was \cite{2024MNRAS.52711756M}, who observed NGC~1068 using \textit{e}-MERLIN and combined those observations with FRAMEx VLA observations to produce highly resolved imaging ($0\farcs12\times0\farcs05$ or $10~{\rm pc}\times4~{\rm pc}$), revealing a new southern component. Upcoming radio telescopes are ideally positioned to probe the missing spatial scales, such as the Square Kilometre Array Observatory \citep{2019arXiv191212699B} or the next generation VLA \citep[ngVLA;][]{2018ASPC..517....3M}, and may provide a wider range of frequencies and deeper sensitivities and at the same time a wide range of spatial scales. For example, utilizing the multi-configuration aspect of the ngVLA, an observer can combine the mid-baseline stations with the spiral array configuration to achieve a resolution of $\sim0\farcs03$ mas at 5 GHz, which spans about 3 pc for a source at a distance of 20 Mpc.\footnote{\url{https://ngect.nrao.edu}}

Lastly, simulations and theoretical modeling may further constrain the radio emission mechanisms. While simulations of collimated jet activity at physical resolutions of subparsec and kiloparsec scales exist \citep{2023ARep...67.1275T,2023ApJ...948...25N}, further high-resolution theoretical models are needed to account for all radio emission mechanisms at the intermediate spatial scales.

In Figure \ref{fig:excess} we observe that above a flux limit of $S_{\rm VLBA}\approx0.35~{\rm mJy~bm^{-1}}$, all FRAMEx targets have been recovered with the VLBA at subparsec spatial scales. Below this threshold, we cannot say for sure that there is no emission for the VLBA non-detections in the subparsec spatial regime and we have not completely ruled out an observational bias in these cases. The FRAMEx VLBA observations may simply lack the sensitivity to observe a faint radio core. Radio emission in AGNs has been observed to show a wide range of flux densities at VLBI resolution \citep[for example, spanning at least 3 dex across 468 targets in][]{2017A&A...607A.132H}, and perhaps an increase in integration time could reveal a faint source \citepalias[as we observed in ][]{2022ApJ...936...76S}. Alternative possibilities for the lack of VLBA detections could potentially be due to radio variability on $\sim$parsec spatial scales, instrumentation or data issues (such as missing antennas during the observation or strong RFI).

This work highlights yet again the complexities of the accretion region in and around SMBHs in a volume-complete sample. Most likely, multiple emission mechanisms contribute to the observed radio structures, whether they are pointlike or extended, and high resolution observations from ancillary data across the electromagnetic spectrum combined with high resolution magneto-hydrodynamic simulations and models are necessary to disentangle all of the contributing factors.

\section*{acknowledgments}

The National Radio Astronomy Observatory is a facility of the National Science Foundation operated under cooperative agreement by Associated Universities, Inc. The authors acknowledge use of the Very Long Baseline Array under the US Naval Observatory’s time allocation. This work supports USNO’s ongoing research into the celestial reference frame and geodesy.

\software{CASA \citep{2022PASP..134k4501C}, astropy \citep{astropy:2013,astropy:2018,astropy:2022}, scipy \citep{2020SciPy-NMeth}, astroquery \citep{2019AJ....157...98G}}
\facilities{VLA, VLBA, Swift}

\bibliography{sample701}{}
\bibliographystyle{aasjournal}

\end{document}